\documentclass[twocolumn,showpacs,preprintnumbers,amsmath,amssymb]{revtex4}
\usepackage{dsfont}
\usepackage{cases}
\usepackage{graphicx}
\usepackage{dcolumn}
\usepackage{bm}
\usepackage{amsmath}
\usepackage{hyperref}
\usepackage{cleveref}

\UseRawInputEncoding

\begin{document}
\title{Modulating Entanglement Dynamics of Two V-type Atoms in Dissipative Cavity by Detuning, Weak Measurement and Reversal}
\author{Fuhua Li}%
\author{Jia Wang}
\author{Hong-Mei Zou}
\email{zhmzc1997@hunnu.edu.cn}
\author{Qiying Pan}%
\author{Chenya Liu}
\author{Zhuang Qu}
\affiliation{Synergetic Innovation Center for Quantum Effects and Application, Key Laboratory of Low-dimensional Quantum Structures and Quantum Control of Ministry of Education, School of Physics and Electronics, Hunan Normal University, Changsha, 410081, People's Republic of China.}%

\date{\today}
\begin{abstract}
In this paper, how to modulate entanglement dynamics of two V-type atoms in dissipative cavity by detuning, weak measurement and weak measurement reversal is studied. The analytical solution of this model is obtained by solving Schr\"{o}dinger Equation after diagonalizing Hamiltonian of dissipative cavity. It is discussed in detail how the entanglement dynamics is influenced by cavity-environment coupling, spontaneously generated interference (SGI) parameter, detuning between cavity with environment and weak measurement reversal. The results show that the entanglement dynamics of different initial states obviously depends on coupling, SGI parameter, detuning and reversing measurement strength. The stronger coupling, the smaller SGI parameter, the larger detuning and the bigger reversing measurement strength can all not only protect but also generate the entanglement, and the detuning is more effectively in tne strong coupling regime than the weak measurement reversal, which is more effectively than the SGI parameter. We also give corresponding physical interpretations.

\end{abstract}

\maketitle

\section{Introduction}
Quantum entanglement was considered as a fundamental concept in  quantum physics, which has been as a key resource for quantum computation and quantum information \cite{Nielsen}.
Many new schemes based on quantum entanglement have been proposed in quantum information processing. For example, some theoretical and experimental proposals about quantum teleportation have been  put forward successively \cite{Pirandola,Bouwmeester,Yan,Cacciapuoti,Lipka,Llewellyn}.
Some new techniques, including reverse engineering and geometric phases as well as composite pulses, can improve the speed and robustness of quantum computing processes \cite{Lefforge,Hou,Kang,Torosov,Wu}.
In quantum communication, some new schemes about dense encoding sources and protocols have been constructed \cite{Mattle,Meher,Shaukat,Guo}, and many new methods have been proposed successively in quantum key distribution \cite{Xu F,Li Z,Fossier,Zhang}.
However inevitable coupling between quantum system and its environment will lead to quantum decoherence for open systems, and decoherence usually leads to decay and even sudden death of entanglement \cite{Rijavec,Eberly,Yu,Almeida}. Therefore, it becomes very important to put forward some new schemes to protect quantum entanglement aganist decoherence in open systems.

In the past decades, many effectively regulating methods have been proposed in order to protect quantum entanglement of open systems. For example, the detuning of cavity field or environment \cite{Fang,Golkar},
the feedback and memory effects of non-Markovian environment \cite{Zou,Zou H,Mu}, the dipole-dipole interaction between atoms \cite{Altintas,Fasihi}, the external classical driving \cite{Wang Q,Golkar}, the auxiliary qubits \cite{Behzad}, the quantum Zeno effect \cite{Maniscalco}, the quantum weak measurement and reversal can all effectively protect entanglement of qubits from decoherence \cite{Sun,Kim,Liao,Man,Y,Liu,Xu,Huang}.

In above mentioned studies, most of them focus on qubit systems. However, in recent years, some methods of regulating entanglement in open qubits have also been generalized to qutrits, which has more advantages in quantum information processing. For example, the qubit-qutrit entanglement can improve the efficiency of communication channels \cite{Lanyon}. The entanglement between two qutrits can be preserved by auxiliary qutrits \cite{Faizi,Ahansaz}. The results in Refs.\cite{Metwally,Li G,Parvin} show that a classical driving can validly preserve and regulate quantum entanglement. Moreover, the weak measurement and reversal can actively combat amplitude damping decoherence in a qutrit-qutrit system \cite{YL}. The authors in \cite{Wang} investigated entanglement dynamics of two V-type atoms with dipole-dipole interaction in dissipative cavity. Xing Xiao \textit{et al} found that entanglement can be protected by detuning between atom and cavity \cite{Obada,Tan}.

Motivated by these works, we will investigate entanglement dynamics of two V-type atoms in dissipative cavity under the combination of detuning and weak measurement reversal. Our goal is to understand whether entanglement of two qutrits can been protected and generated by the detuning between cavity and environment, the weak measurement and its reversal. 
We discover that the detuning and the weak measurement reversal can all protect very effectively the maximal entangled state, and promote the generation of entanglement in the product state. It provides some references for theoretical and experimental researches of opening qutrit systems.

In this paper, we propose a scheme to modulate the entanglement dynamics of two identical V-type atoms in a dissipative single-mode cavity by adjusting the detuning between cavity and environment and using the weak measurement reversal. The results show that the stronger coupling, the smaller SGI parameter, the larger detuning and the bigger reversing measurement strength can all not only protect but also generate the entanglement, and the detuning is more effectively in tne strong coupling regime than the weak measurement reversal, which is more effectively than the SGI parameter.

The outline of paper is following. In Section II, we introduce physical model and obtain the density operator of two V-type atoms. In Section III, we give the entanglement negativity of two atoms under weak measurement and its reversal. Results and discussions are provided in Section IV. Finally, we give a brief summary in Section V.

\section{ Physical model}
\subsection{ Schr\"{o}dinger equation}
\begin{figure}[tbp]
    \includegraphics[width=7cm,height=4cm]{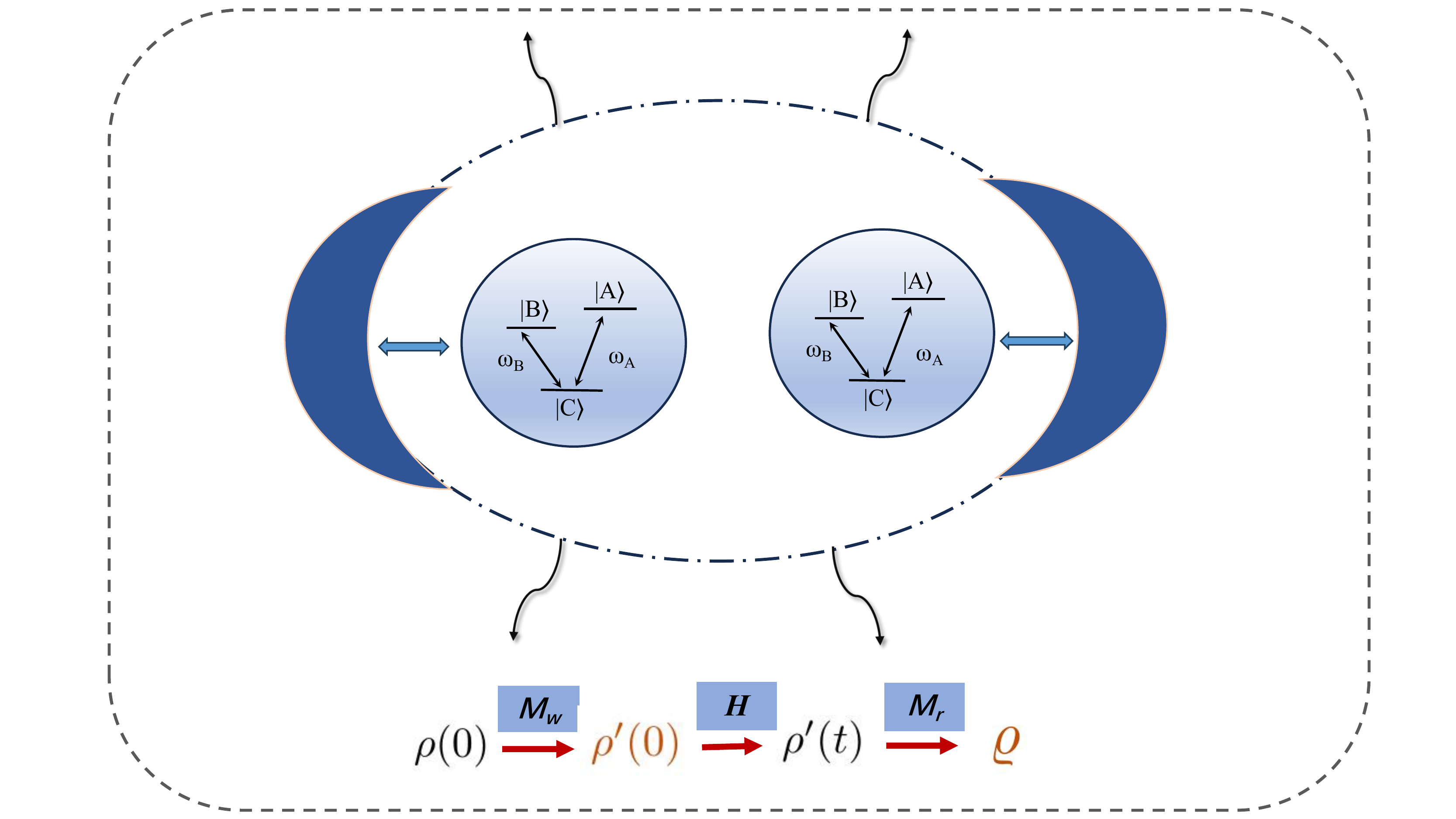}
	\caption{(Color online) Two identical V-type atoms interact with a dissipative cavity and each atom is subjected to a prior weak measurement and a post weak measurement.  The atomic transition frequencies from the ground state $|C\rangle$ to the excited states $|A\rangle$ and $|B\rangle$ are $\omega _{A}$ and $\omega _{B}$, respectively.}
	\label{fig:1}
\end{figure}

We structure a combined system of two identical V-type atoms in a dissipative cavity, where each qutrit is subjected to a prior weak measurement and a post weak measurement reversal, and the cavity is characterized by a single-mode field and its reservoir is described by a set of continuum harmonic oscillators, as depicted in Fig.1. The excited states $|A\rangle$ and $|B\rangle$ can spontaneously decay into the ground state $|C\rangle$ with the transition frequencies $\omega _{A}$ and $\omega _{B}$, respectively. Ignoring the dipole-dipole interaction between the atoms and under the rotating-wave approximation\cite{Agarwal}, the total Hamiltonian($\hbar =1$) is given by
\begin{equation} \label{EB201}
\begin{aligned}
\hat{H}=\hat{H}_{s}+\hat{H}_{s c}+\hat{H}_{cr}
\end{aligned}
\end{equation}
here
\begin{equation}\label{EB202}
\hat{H}_{s}=\sum_{l=1}^{2} \sum_{m=A, B} \omega_{m} \hat{\sigma}_{m}^{l+} \hat{\sigma}_{m}^{l-}
\end{equation}
is the atomic Hamiltonian and $\hat{\sigma}_{m}^{l+ }$ and $\hat{\sigma}_{m}^{l- }$ $(m = A,B)$ are the raising and lowering operators of the $m$th excited state of the $l$th atom, and $\hat{\sigma}_{m}^{l+}=|m\rangle_{l} \left\langle C\right|$ and $\hat{\sigma}_{m}^{l-}=|C\rangle_{l}\langle m|$ \cite{Behzadi,Li} are the transition operators with the frequency $\omega _{m}$.  The interaction Hamiltonian $\hat{H}_{sc}$ between the atoms with their cavity is
\begin{equation}\label{EB203}
\hat{H}_{sc}=\sum_{l=1}^{2} \sum_{m=A, B} g_{m} \hat{\sigma}_{m}^{l+}\hat{a} + g_{m} ^{\ast }\hat{\sigma}_{m}^{l-}\hat{a}^{\dagger }
\end{equation}
where $\hat{a}^{\dagger }$ and $\hat{a}$ are the creation and annihilation operators of single-mode cavity respectively, and the coupling strength between the $m$th excited state and the cavity is given by $g_{m}$. The Hamiltonian of dissipative cavity is
\begin{equation}\label{EB204}
\begin{split}
\hat{H}_{cr}=\omega_{c}\hat{a}^{\dagger }\hat{a}+\int_{0}^{\infty } \eta \hat{B}^{\dagger  }{(\eta) }\hat{B }(\eta)d\eta\\
+\int_{0}^{\infty }
\left \{G(\eta)\hat{a}^{\dagger }\hat{B}(\eta)+H.C. \right \}d(\eta)
\end{split}
\end{equation}
where $\omega_{c}$ is the cavity eigenfrequency, $\hat{B}^{\dagger  }(\eta)$ and $\hat{B}(\eta)$ are the creation and annihilation operators of reservoir in the mode $\eta$ with the commutation relation $\left [\hat{B }(\eta), \hat{B}^{\dagger  }(\eta{}' )  \right ] =\delta \left (\eta-\eta{}'   \right )$. $ {G }(\eta)=\sqrt{\kappa /\pi } $ is the coupling coefficient of cavity-reservoir, $\kappa$ is the decay rate of cavity and H.C. expresses the Hermitian conjugation.

In order to diagonalize Hamiltonian $\hat{H}_{cr}$, we introduce a new annihilation operator $\hat{A}(\omega )$\cite{Nourmandipour} according to Fano theorem, which is defined as
\begin{equation}\label{EB205}
\hat{A}(\omega )=\alpha (\omega )\hat{a}+\int \beta (\omega ,\eta )\hat{B}(\eta )d\eta.
\end{equation}
and
\begin{equation}\label{EB206}
\left [\hat{A} (\omega), \hat{A}^{\dagger} ({\omega}') \right ] =\delta (\omega -{\omega}')
\end{equation}

where
\begin{equation}\label{EB207}
\alpha (\omega )=\frac{\sqrt{\kappa/ \pi}}{ \omega-\omega_{c}  +i\kappa }
\end{equation}

\begin{equation}\label{EB208}
\beta(\omega,\eta)=\sqrt{\kappa/\pi}\alpha(\omega)P\frac{1}{\omega-\eta}+\frac{\omega-\omega_{c}}{\sqrt{\kappa/\pi}}\alpha(\omega)\delta(\omega-\eta)
\end{equation}
where $P$ is the principal value\cite{Nourmandipour}. Then we can get the linear combination about $\hat{a}$
\begin{equation}\label{EB209}
\hat{a}=\int_{0}^{\infty}\alpha^\ast \left(\omega\right)\hat{A}\left(\omega\right)d\omega
\end{equation}
 Then Eq.(4) becomes
\begin{equation}\label{EB2010}
\hat{H}_{cr}=\int \omega \hat{A}^{\dagger} (\omega )\hat{A}(\omega )d\omega
\end{equation}
and
\begin{equation}\label{EB2011}
\begin{bmatrix}\hat{A}(\omega),\hat{H}_{cr}\end{bmatrix}=\omega\hat{A}\left(\omega\right)
\end{equation}

If the cavity and its reservoir together are regarded as the atomic environment, the total Hamiltonian Eq.(1) can be rewritten as
\begin{equation}\label{EB2012}
\hat{H}=\hat{H}_{0}+\hat{H}_{I}
\end{equation}
where $\hat{H}_{0}$ is the free Hamiltonian, i.e.
\begin{equation}\label{EB2013}
\hat{H}_{0}=\sum_{l=1}^{2} \sum_{m=A, B} \omega_{m} \hat{\sigma}_{m}^{l+} \hat{\sigma}_{m}^{l-}+\int \omega \hat{A}^{\dagger} (\omega )\hat{A}(\omega )d\omega
\end{equation}
and $\hat{H}_{I}$ is the interaction Hamiltonian
\begin{equation}\label{EB2014}
\hat{H}_{I}=\sum_{l=1}^{2} \sum_{m=A, B}\int \left \{    g_{m}\alpha ^{\ast }(\omega )
 \hat{\sigma}_{m}^{l+}\hat{A}(\omega) + H.C.\right \}d\omega
\end{equation}
In the interaction picture, the Schr\"{o}dinger equation is
\begin{equation}\label{EB2015}
i\frac{\mathrm{d}}{\mathrm{d}t}|\psi(t)\rangle=\hat{H}_{\mathrm{int}}(t)|\psi(t)\rangle
\end{equation}
where $\hat{H}_{\mathrm{int}}(t)=\mathrm{e}^{i\hat{H}_0t}\hat{H}_I\mathrm{e}^{-i\hat{H}_0t}$.

Let the total system has only one excitation and is initial in
\begin{equation}\label{EB2016}
|\psi(0)\rangle=(C_{2}^{A}\left(0\right)|C_{1}A_{2}\rangle+C_{1}^{B}\left(0\right)|B_{1}C_{2}\rangle)\otimes|0\rangle_{R}
\end{equation}
where $|C_{1}A_{2}\rangle$ represents that the first atom is in the ground state $|C\rangle$ and the other is in the excited state $|A\rangle$, $|B_{1}C_{2}\rangle$ indicates that the first atom is in the excited state $|B\rangle$ and the other is in the ground state $|C\rangle$. $|C_{2}^{A}(0)|^{2}+|C_{1}^{B}(0)|^{2}=1$ and  $|0\rangle_{_R}=\hat{A}(\omega)|1_{\omega^{\prime}}\rangle\delta\left(\omega-\omega^{\prime}\right)$. $|0\rangle_{R}$ denotes that the environment is in the vacuum state and $|1_{\omega}\rangle$ indicates that the environment has only one excitation in the mode with frequency $\omega$.

We may write the time evolution state as
\begin{equation}\label{EB2017}
\begin{aligned}
|\psi(t)\rangle & =\left(C_{1}^{A}(t)|A_{1}C_{2}\rangle+C_{1}^{B}(t)|B_{1}C_{2}\rangle\right)_{S}\otimes|0\rangle_{\mathrm{R}}  \\
&\left.+\left(\left.C_{2}^{A}(t)\right|C_{1}A_{2}\right>+\left.C_{2}^{B}(t)\right|C_{1}B_{2}\rangle\right)_{S}\otimes|0\rangle_{\mathrm{R}} \\
&+\int C_{\omega}(t)|C_{1}C_{2}\rangle_{S}|1_{\omega}\rangle_{\mathrm{R}}\mathrm{d}\omega
\end{aligned}
\end{equation}
By substituting Eq.(17) into Eq.(15), we can obtain the integro-differential equation of probability amplitude
\begin{equation}\label{EB2018}
\begin{split}
\frac{dC_{l}^{m}\left(t\right)}{dt}=-\sum_{n=A,B}\int_{0}^{t}f_{mn}\left(t-t^{\prime}\right)\sum_{j=1}^{2}C_{j}^{n}\left(t^{\prime}\right)dt^{\prime}
\end{split}
\end{equation}
where $m=A,B$, the correlation function $f_{mn}(t-t^{\prime})$ is relevant with the spectral density $J_{mn}(\omega)$ of environment by
\begin{equation}\label{EB2019}
\begin{split}
f_{mn}(t-t^{\prime})=\int d\omega J_{mn}(\omega)e^{i(\omega_m-\omega)t-i(\omega_n-\omega)t^{\prime}}
\end{split}
\end{equation}

Let the spectral density has Lorentzian form as
\begin{equation}\label{EB2020}
J_{mn}(\omega)=\frac1{2\pi}\frac{\gamma_{mn}\kappa^2}{\left(\omega-\omega_c-\delta \right)^2+\kappa^2}\quad\quad
\end{equation}
where $\delta$ is the detuning between the cavity eigenfrequency and the central frequency of environment, and
$\gamma_{mn}=\frac{2g_mg_n^*}\kappa$ is the relaxation rate of excited states and
\begin{equation}\label{EB2021}
\gamma_{mm}=\gamma_m
\end{equation}
\begin{equation}\label{EB2022}
\gamma_{mn}=\sqrt{\gamma_m\gamma_n}\theta,m\neq n,|\theta|\le1
\end{equation}
where $\theta$ is defined as the SGI (spontaneously generated interference) parameter between the two decay channels $|A\rangle \to  |C\rangle$ and $|B\rangle \to |C\rangle$ of each atom, which $\theta$ depends on the angle between two dipole moments of transition. $\theta=1$ indicates that the two dipole moments are parallel, which means that there is the strongest SGI between the two decay channels. When $\theta=0$, the two dipole moments are perpendicular to each other and there is no SGI between the two decay channels \cite{Wang}.

Without loss of generality, we assume that the two upper atomic states are degenerated and the atomic transition frequencies are resonant with the cavity eigenfrequency, i.e. $\omega _{A}=\omega _{B}=\omega _{c}$, $\gamma  _{A}=\gamma  _{B}=\gamma  _{0}$ and $\gamma  _{0}$ is the decay coefficient of the atomic excited state as well as
$\gamma  _{A B}=\gamma  _{BA}=\gamma  _{0}\theta$. $\kappa>>2\gamma_{0}$ indicates a weak (Markovian) coupling between cavity and environment while $\kappa<<2\gamma_{0}$ means a strong (non-Markovian) coupling between cavity and environment \cite{Wang,Bellomo}. Solving the integrodifferential Eq.(18), the probability amplitudes can be expressed as (see Appendix)

\begin{equation}
C_{l}^{A}(t) =\frac{1}{2}(C_{l}^{+}(t)+C_{l}^{-}(t))
\end{equation}

\begin{equation}
C_{l}^{{B}}(t)=\frac{1}{2}(C_{l}^{+}(t)-C_{l}^{-}(t))
\end{equation}
here
\begin{equation}
C_{1}^{\pm  }(t)=G_ {\pm }(t)C_{1}^{\pm  }(0)-\frac{1-2G_{\pm }(t)}{2}[C_{2}^{\pm }(0)-C_{1}^{\pm }(0)]
\end{equation}
\begin{equation}
C_{2}^{\pm  }(t)=G_ {\pm }(t)C_{2}^{\pm  }(0)-\frac{1-2G_{\pm }(t)}{2}[C_{1}^{\pm }(0)-C_{2}^{\pm }(0)]
\end{equation}
where
\begin{equation}
\mathcal{G_{\pm} }\left(t\right) =e^{-(\kappa-i\delta ) t/2}\bigg(\cosh\left(\frac{d_{\pm} t}{2}\right)+\frac{\kappa-i\delta }{d_{\pm}}\sinh\left(\frac{d_{\pm}t}{2}\right)\bigg)
\end{equation}
and
\begin{equation}
d_{\pm } =\sqrt{(\kappa-i\delta )^{2} -4\gamma _{0}\kappa (\theta \pm 1) }
\end{equation}

According to Eqs.(23-26), we know that all probability amplitudes can be written as the following simple forms
\begin{equation}
\begin{split}
C_{1}^{A}(t)=G_{1}(t)C_{1}^{A}(0)+G_{2}(t)C_{2}^{A}(0)\\
+G_{3}(t)[C_{1}^{B}(0)+C_{2}^{B}(0)]
\end{split}
\end{equation}

\begin{equation}
\begin{split}
C_{1}^{B}(t)=G_{1}(t)C_{1}^{B}(0)+G_{2}(t)C_{2}^{B}(0)\\
+G_{3}(t)[C_{1}^{A}(0)+C_{2}^{A}(0)]
\end{split}
\end{equation}

\begin{equation}
\begin{split}
C_{2}^{A}(t)=G_{1}(t)C_{2}^{A}(0)+G_{2}(t)C_{1}^{A}(0)\\
+G_{3}(t)\left[C_{1}^{B}(0)+C_{2}^{B}(0)\right]
\end{split}
\end{equation}

\begin{equation}
\begin{split}
C_{2}^{B}(t)=G_{1}(t)C_{2}^{B}(0)+G_{2}(t)C_{1}^{B}(0)\\
+G_{3}(t)\left[C_{1}^{A}(0)+C_{2}^{A}(0)\right]
\end{split}
\end{equation}

where
\begin{equation}
G_{1}(t)=\frac{G_{+ }(t)+G_{- }(t)+2}{4}
\end{equation}

\begin{equation}
G_{2}(t)=\frac{G_{+ }(t)+G_{- }(t)-2}{4}
\end{equation}

\begin{equation}
G_{3}(t)=\frac{G_{+ }(t)-G_{- }(t)}{4}
\end{equation}

\subsection{ Weak measurement and weak measurement reversal}
In this subsection, we introduce an optimal scheme to combat decoherence: for each identical V-type atom, a prior weak measurement with strength $p$  is performed before the atom-cavity interaction, then a post weak measurement reversal with strength $p_{r}$ is carried out after the atom-cavity interaction,  as shown in Fig.1. For the two V-type atoms, the weak measurement and weak measurement reversal are non-unitary operators, which can be written respectively as 
\begin{equation}
\begin{split}
M_w(p_{i},q_{i})=\left(\begin{array}{ccc}1&0&0\\0&\sqrt{1-p_{1} }&0\\0&0&\sqrt{1-q_{1} }\end{array}\right)\\
\otimes \left(\begin{array}{ccc}1&0&0\\0&\sqrt{1-p_{2} }&0\\0&0&\sqrt{1-q_{2} }\end{array}\right)
\end{split}
\end{equation}
and
\begin{equation}
\begin{split}
M_r(p_{r_{i}},q_{r_{i}})  =\left(\begin{array}{ccc}\sqrt{(1-p_{r_{1}}  )(1-q_{r_{1}} )}&0&0\\0&\sqrt{1-q_{r_{1}} }&0\\0&0&\sqrt{1-p_{r_{1}} }\end{array}\right) \\ \otimes \left(\begin{array}{ccc}\sqrt{(1-p_{r_{2}}  )(1-q_{r_{2}} )}&0&0\\0&\sqrt{1-q_{r_{2}} }&0\\0&0&\sqrt{1-p_{r_{2}} }\end{array}\right)
\end{split}
\end{equation}
where $M_w(p_{i},q_{i})$ is the weak measurement operator, $p_{i}$ and $q_{i}$ are weak measurement strengths of the $i$th atom ($i$=1,2). In the same presentation, $M_r(p_{r_{i}}, q_{r_{i}})$ is the reversing measurement operator, $p_{r_{i}}$ and $q_{r_{i}}$ are the reversing measurement strengths. For simplicity, we assume that $p_{1}= q_{1}=p_{2}= q_{2} = p$ and $p_{r_{1}}=q_{r_{1}}=p_{r_{2}}=q_{r_{2}}=p_{r}$. As the matrix is non-unitary, the probability of successful reversal will always be less than 1.

\subsection{ Density matrix of two V-type atoms}
Inserting Eqs.(29-32) into Eq.(17), we can get the density operator of the total system, i.e. $\rho_{tot}(t)=\left | \psi(t)  \right \rangle \left\langle\psi(t)\right| $. Then tracing the environmental freedom degree $\rho(t)=Tr_{E}(\rho_{tot}(t))$, we can obtain the reduced density operator of two V-type atoms in the basis
$\left \{  \left | C_1 C_2 \right \rangle,\left | C_1 B_2 \right \rangle,\left | C_1 A_2 \right \rangle,\left | B_1 C_2 \right \rangle,
\left | B_1B_2 \right \rangle,\left | B_1 A_2 \right \rangle,\right .$\\
$\left.\left | A_1 C_2 \right \rangle,\left | A_1 B_2 \right \rangle,\left |A_1 A_2 \right \rangle\right \}$.

Since the non-zero elements of $\rho(0)$ are following
\begin{equation}
\begin{split}
\rho _{33}(0)=|C_2^A(0)|^2,  \rho _{44}(0)=|C_1^B(0)|^2 \\
\rho _{34}(0)=\rho _{43}^{*}(0)=C_{2}^{A}(0)C_{1}^{B*}(0)
\end{split}
\end{equation}
After the weak measurement Eq.(36), $\rho^{\prime}(0)=M_{\omega } \rho(0)M_{\omega}^{\dagger }$, the non-zero element of $\rho'(0)$ are
\begin{equation}
\begin{split}
\rho _{33}'(0)=|C_2^A(0)|^2(1-p)/C_{1},\\
\rho _{44}'(0)=\rho _{44}(0)(1-p)/C_{1},\\
\rho_{34}'(0)=\rho_{43}'^{*}(0)=C_{2}^{A}(0)C_{1}^{B*}(0)(1-p)/C_{1}.
\end{split}
\end{equation}
where $C_{1}=1-p$ is the normalization parameter. The non-zero element of $\rho(t)$ are
\begin{equation}
\begin{split}
\rho _{11}=|C_{\omega }(t)|^2, \rho _{22}=|C_2^{B}(t)|^2,\\
\rho _{33}=|C_2^{A}(t)|^2, \rho _{44}=|C_1^B(t)|^2,\\
\rho _{23}=\rho _{32}^{*}=C_{2}^{B}(t)C_{2}^{A*}(t), \\
\rho _{24}=\rho _{42}^{*}=C_{2}^{B}(t)C_{1}^{B*}(t), \\
\rho _{34}=\rho _{43}^{*}=C_{2}^{A}(t)C_{1}^{B*}(t), \\
\rho _{27}=\rho _{72}^{*}=C_{2}^{B}(t)C_{1}^{A*}(t), \\
\rho _{37}=\rho _{73}^{*}=C_{2}^{A}(t)C_{1}^{A*}(t), \\
\rho _{47}=\rho _{74}^{*}=C_{1}^{B}(t)C_{1}^{A*}(t), \\
\rho _{77}=|C_1^A(t)|^2,
\end{split}
\end{equation}
where$|C_{\omega}(t)|^2=1-|C_1^A(t)|^2-|C_1^B(t)|^2-|C_2^A(t)|^2-|C_2^B(t)|^2$.

We can get $\rho' (t)$ from Eqs.(39-40) and then perform the quantum measurement reversal on $\rho' (t)$, i.e. $\varrho =M_{r } \rho' (t)M_{r}^{\dagger }$. The non-zero elements of $\varrho$ are
\begin{equation}
\begin{split}
\varrho _{11}=|C_{\omega }^{'}(t)|^2(1-p_{r})^{4}/C_{2},\\
\varrho  _{22}=|C_2^{B'}(t)|^2(1-p_{r})^{3}/C_{2}, \\
\varrho  _{33}=|C_2^{A'}(t)|^2(1-p_{r})^{3}/C_{2}, \\
\varrho _{44}=|C_1^{B'}(t)|^2(1-p_{r})^{3}/C_{2},\\
\varrho  _{77}=|C_1^{A'}(t)|^2(1-p_{r})^{3}/C_{2}, \\
\varrho _{23}=\varrho _{32}^{*}=C_{2}^{B'}(t)C_{2}^{A'*}(t)(1-p_{r})^{3}/C_{2},\\
\varrho  _{24}=\varrho  _{42}^{*}=C_{2}^{B'}(t)C_{1}^{B'*}(t)(1-p_{r})^{3}/C_{2},\\
\varrho  _{34}=\varrho  _{43}^{*}=C_{2}^{A'}(t)C_{1}^{B'*}(t)(1-p_{r})^{3}/C_{2},\\
\varrho  _{27}=\varrho _{72}^{*}=C_{2}^{B'}(t)C_{1}^{A'*}(t)(1-p_{r})^{3}/C_{2}, \\
\varrho  _{37}=\varrho _{73}^{*}=C_{2}^{A'}(t)C_{1}^{A'*}(t)(1-p_{r})^{3}/C_{2}, \\
\varrho  _{47}=\varrho  _{74}^{*}=C_{1}^{B'}(t)C_{1}^{A'*}(t)(1-p_{r})^{3}/C_{2}.
\end{split}
\end{equation}
where $C_{2}=|C_{\omega}^{'}(t)|^2(1-p_{r})^{4}+(|C_2^{B'}(t)|^2+|C_2^{A'}(t)|^2+|C_1^{B'}(t)|^2+|C_1^{A'}(t)|^2)(1-p_{r})^{3}$
is also the normalization parameter.

\section{Entanglement negativity }

There are many measurement methods to quantify quantum entanglement, for instance, the entanglement negativity \cite{Vidal}, the concurrence entanglement \cite{Wootters}, the relation of entropy entanglement \cite{Wong} and the partial entropy entanglement \cite{Wen}. In order to study the entanglement dynamics of two qutrits under the weak measurement and its reversal, we use the entanglement negativity, which is defined as \cite{Wootters}
\begin{equation}
\mathcal{N}(\varrho )=-2\sum_i\lambda_i
\end{equation}
where $\lambda_i$ is the negative eigenvalue of partial transposition matrix $\varrho ^{T_{1}}(t)$ and $0\leq\mathcal{N}(\varrho \leq1$, where ${N}(\varrho )=0$ means that the two atoms are disentangled, and ${N}(\varrho )=1$ indicates that they are maximal entangled.

According to $\langle i_{A},j_{B}|\rho^{T_{1}}|k_{A},l_{B}\rangle\equiv\langle k_{A},j_{B}|\rho|i_{A},l_{B}\rangle$, we can obtain the partial transposition matrix 
\begin{equation}
\varrho ^{T_{1}}(t)=\begin{pmatrix}\varrho  _{11}&0&0&0&\varrho  _{42}&\varrho_{43}&0&\varrho _{72}&\varrho _{73}\\0&\varrho  _{22}&\varrho _{23}&0&0&0&0&0&0\\0&\varrho  _{32}&\varrho  _{33}&0&0&0&0&0&0\\0&0&0&\varrho  _{44}&0&0&\varrho  _{74}&0&0\\\varrho  _{24}&0&0&0&0&0&0&0&0\\\varrho  _{34}&0&0&0&0&0&0&0&0\\0&0&0&0&\varrho  _{47}&0&\varrho   _{77}&0&0\\\varrho  _{27}&0&0&0&0&0&0&0&0\\\varrho  _{37}&0&0&0&0&0&0&0&0\end{pmatrix}
\end{equation}

As the weak measurement reversal is a non-unitary operator, its success probability is naturally less than 1. Under the optimal weak measurement and weak measurement reversal, for $p = 0$, we can get the corresponding success probability
\begin{equation}
\begin{split}
P=|C_{\omega}^{'}(t)|^2(1-p_{r})^{4}+(|C_2^{B'}(t)|^2+|C_2^{A'}(t)|^2\\
+|C_1^{B'}(t)|^2+|C_1^{A'}(t)|^2)(1-p_{r})^{3}
\end{split}
\end{equation}
It shows that $P\to 0$ when $p_{r}\to 1$.

\section{Results and Discussion}

We will discuss the entanglement dynamics in two different cases: with resonance and with detuning. 

\subsection{ Entanglement dynamics with resonance($\delta=0$)}
In this subsection, we will analyze the effects of cavity-environment coupling, SGI parameter, weak measurement strength and reversing measurement strength on entanglement negativity when the two V-type atoms are in different initial states.

\begin{figure}[tbp]
    \includegraphics[width=4cm,height=3.5cm]{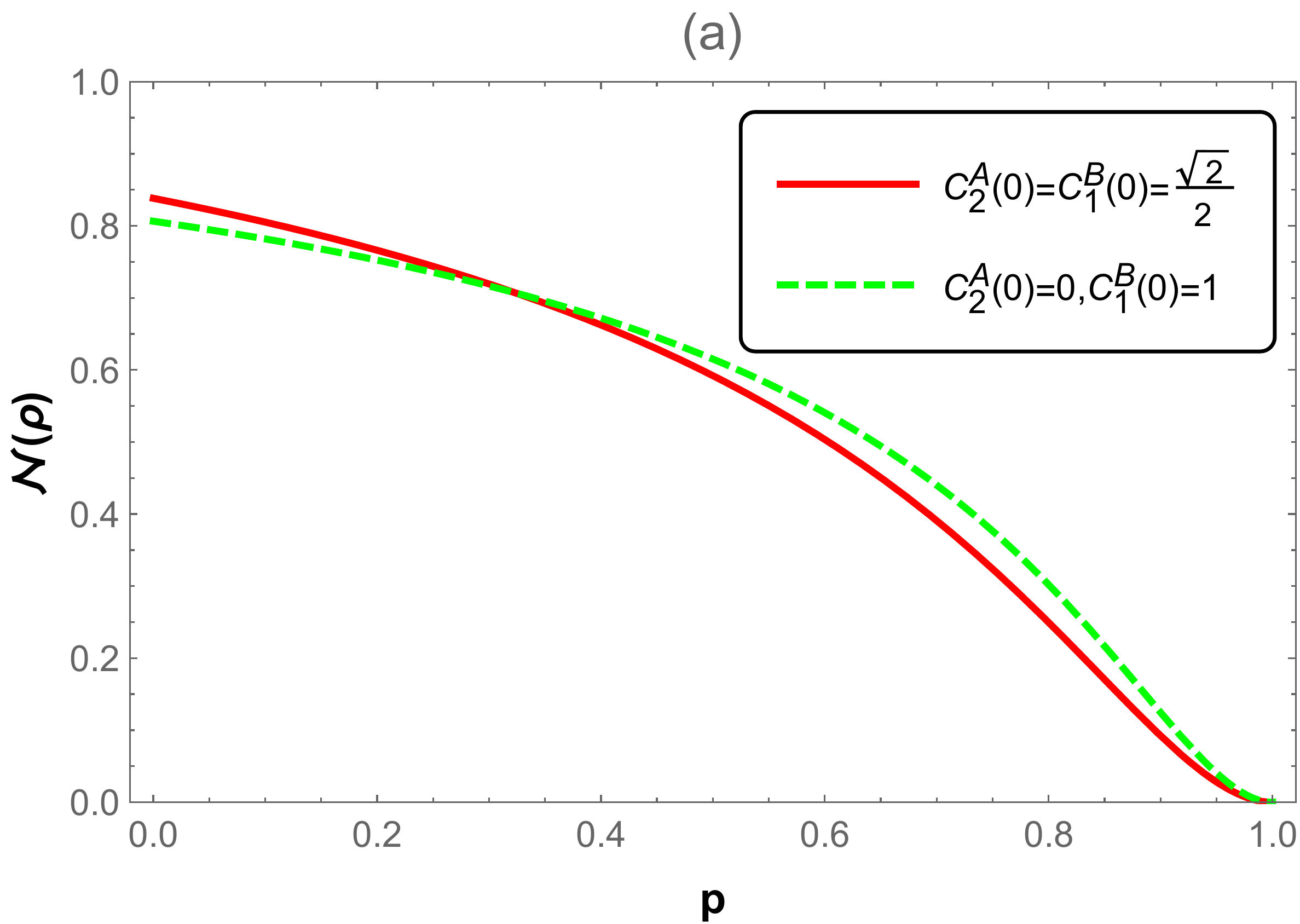}
    \includegraphics[width=4cm,height=3.5cm]{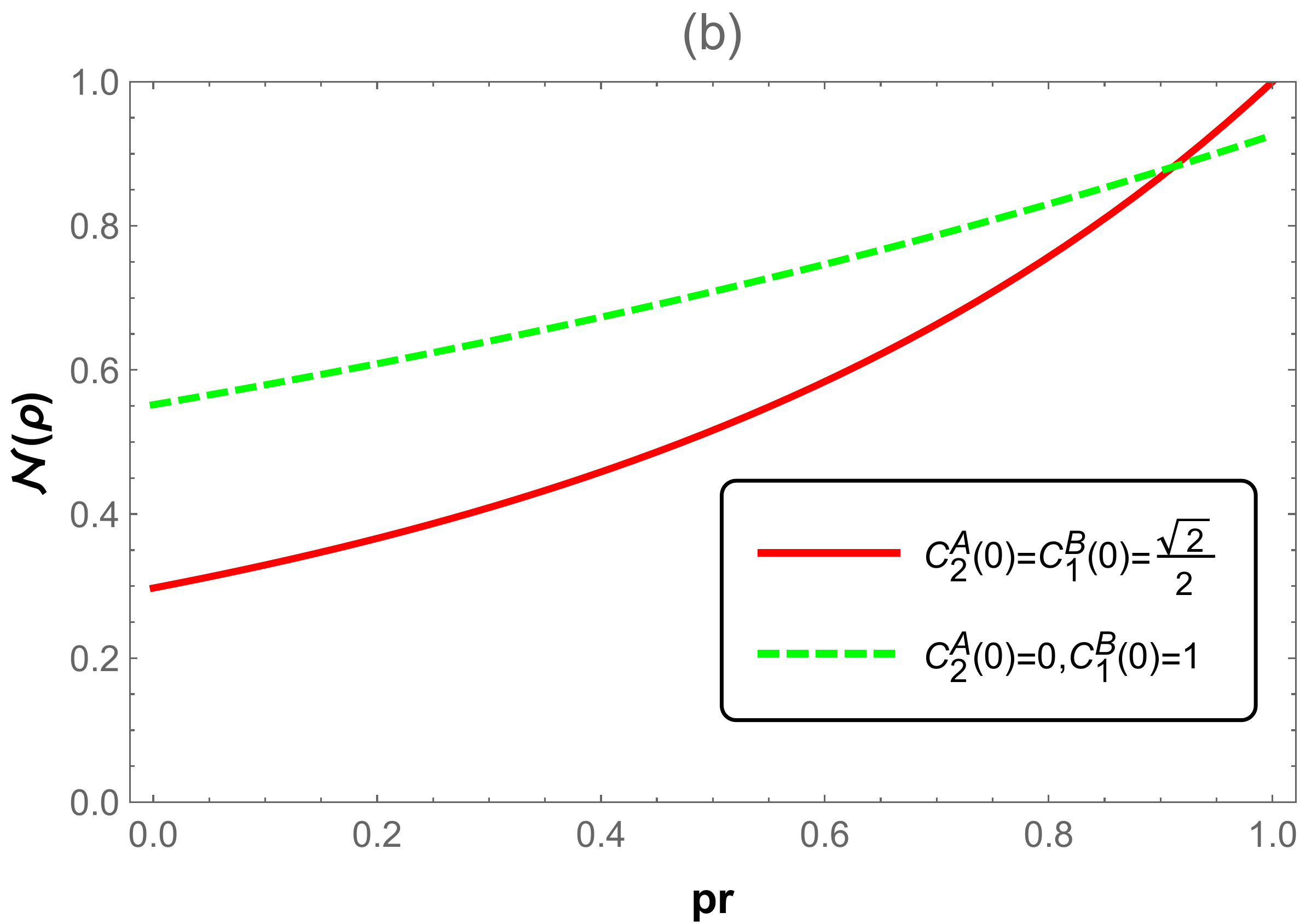}
	\caption{(Color online) Negativity as a function
 of $p$ and $p_{r}$ in the strong coupling regime ($\kappa=0.1\gamma_{0}$) and with the initial states $|\psi(0)\rangle=\frac{\sqrt{2}}{2}(|C_{1}A_{2}\rangle+|B_{1}C_{2}\rangle)_{S}\otimes|0\rangle_{R}$ (red solid line) and $|\psi(0)\rangle=|B_{1}C_{2}\rangle_{S}\otimes|0\rangle_{R}$ (green dashed line). (a)$p_{r}=0$, (b)$p=0$.
The other parameters are $\theta =1$, $\gamma_{0}=1$, $\gamma_{0}t=10$ and $\delta=0$.}
	\label{fig:2}
\end{figure}

In Fig.2, we describe the negativity as a function of parameters $p$ and $p_{r}$ in the strong coupling regime. The initial states are maximal entangled with the coefficients $C_{2}^{A}(0)=C_{1}^{B}(0)=\frac{\sqrt{2}}{2}$ and disentangled with $C_{2}^{A}(0)=0$ and  $C_{1}^{B}(0)=1$, respectively. In Fig.2a, we notice that, when $\gamma_{0}t=10$, the negativity will all decrease to 0 from a bigger value with $p$ increasing for the two different initial states. However, Fig.2b tells us that the negativity always grows from a value with $p_{r}$ increasing. Hence, the entanglement will reduce with $p$ increasing when $p_{r}=0$ while it will grow with $p_{r}$ increasing when $p=0$. Namely, the entanglement can be proteced by a larger $p_{r}$ but will be destroyed by a larger $p$.

\begin{figure}[tbp]
    \includegraphics[width=4cm,height=3.5cm]{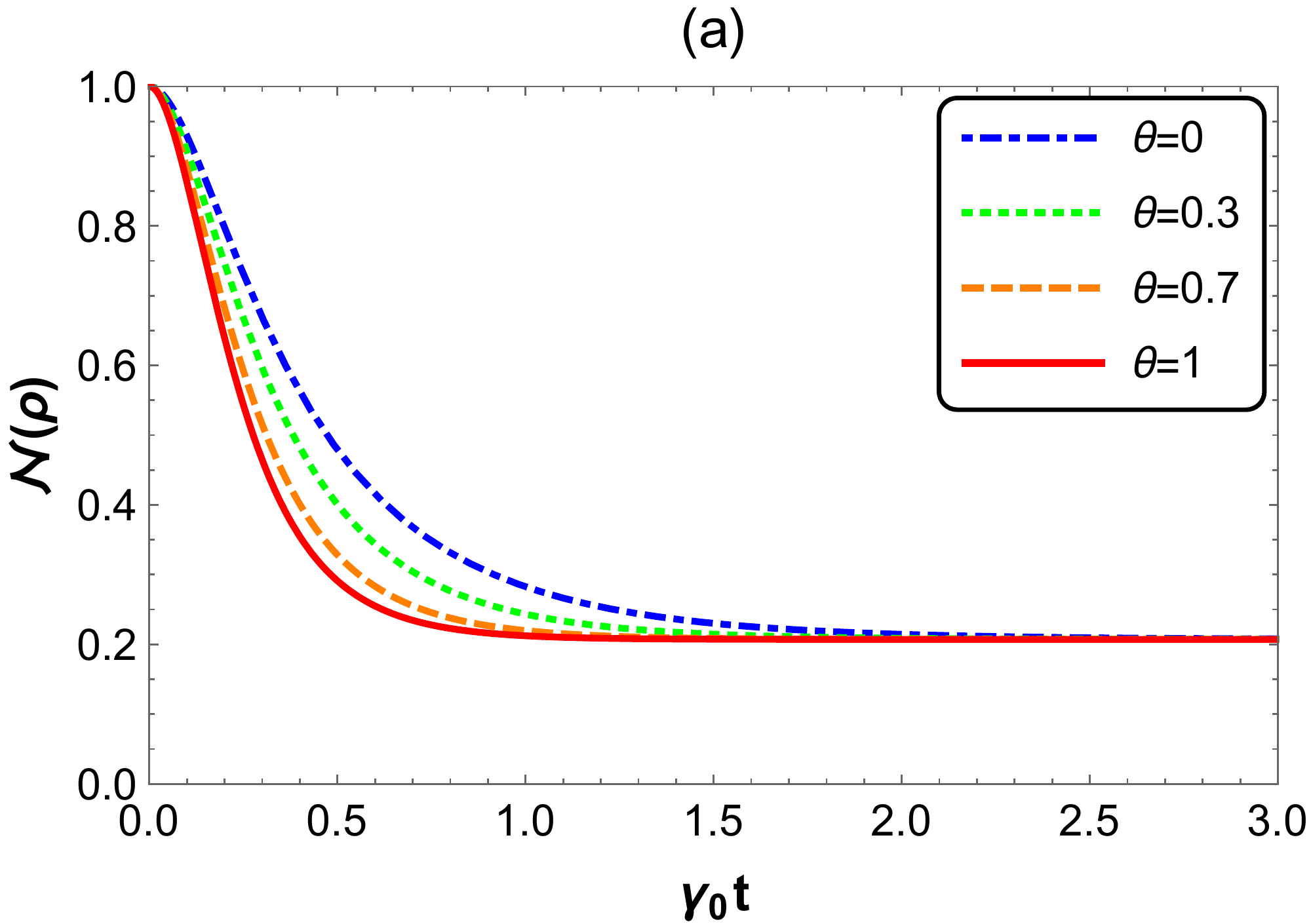}
    \includegraphics[width=4cm,height=3.5cm]{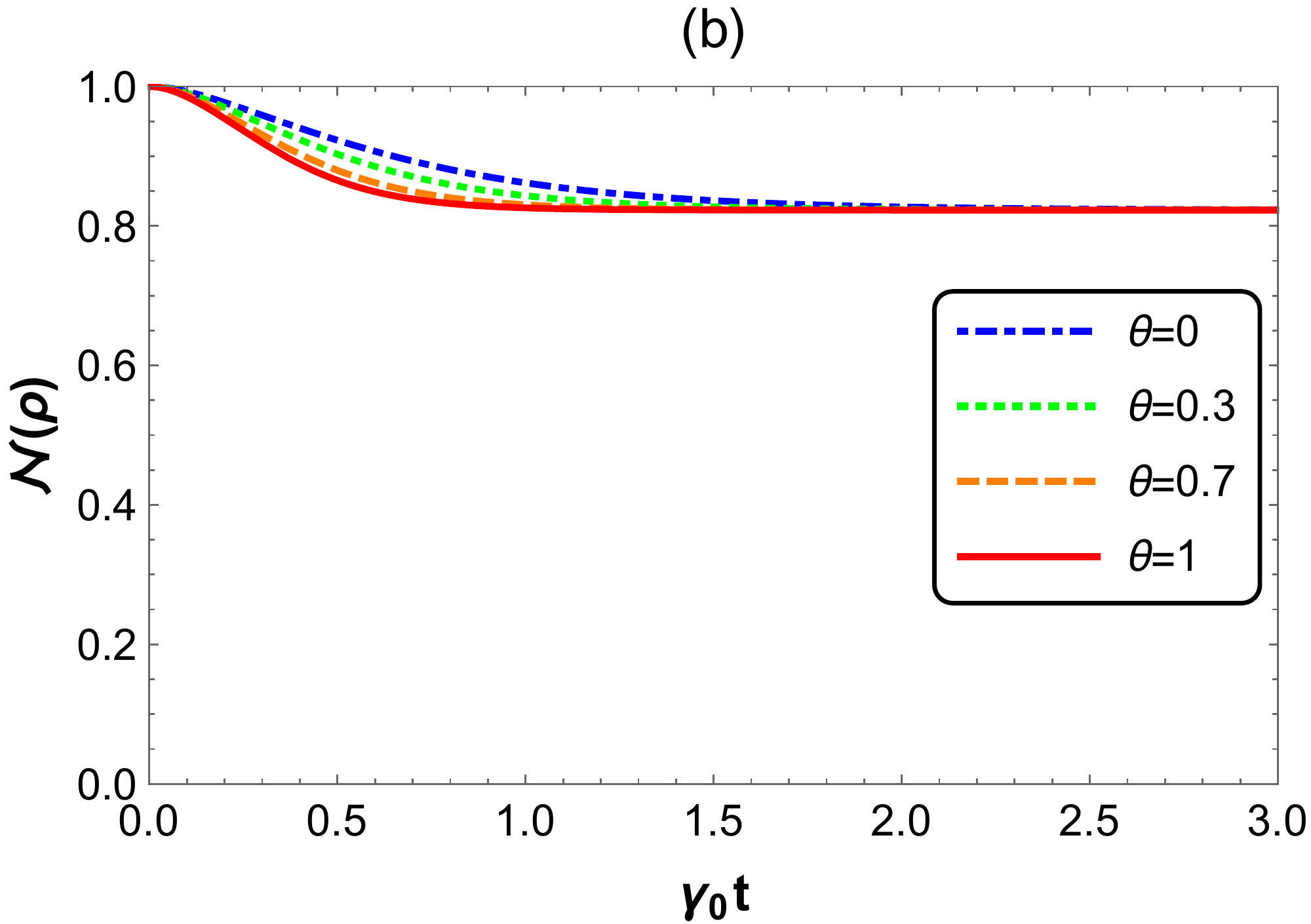}
    \includegraphics[width=4cm,height=3.5cm]{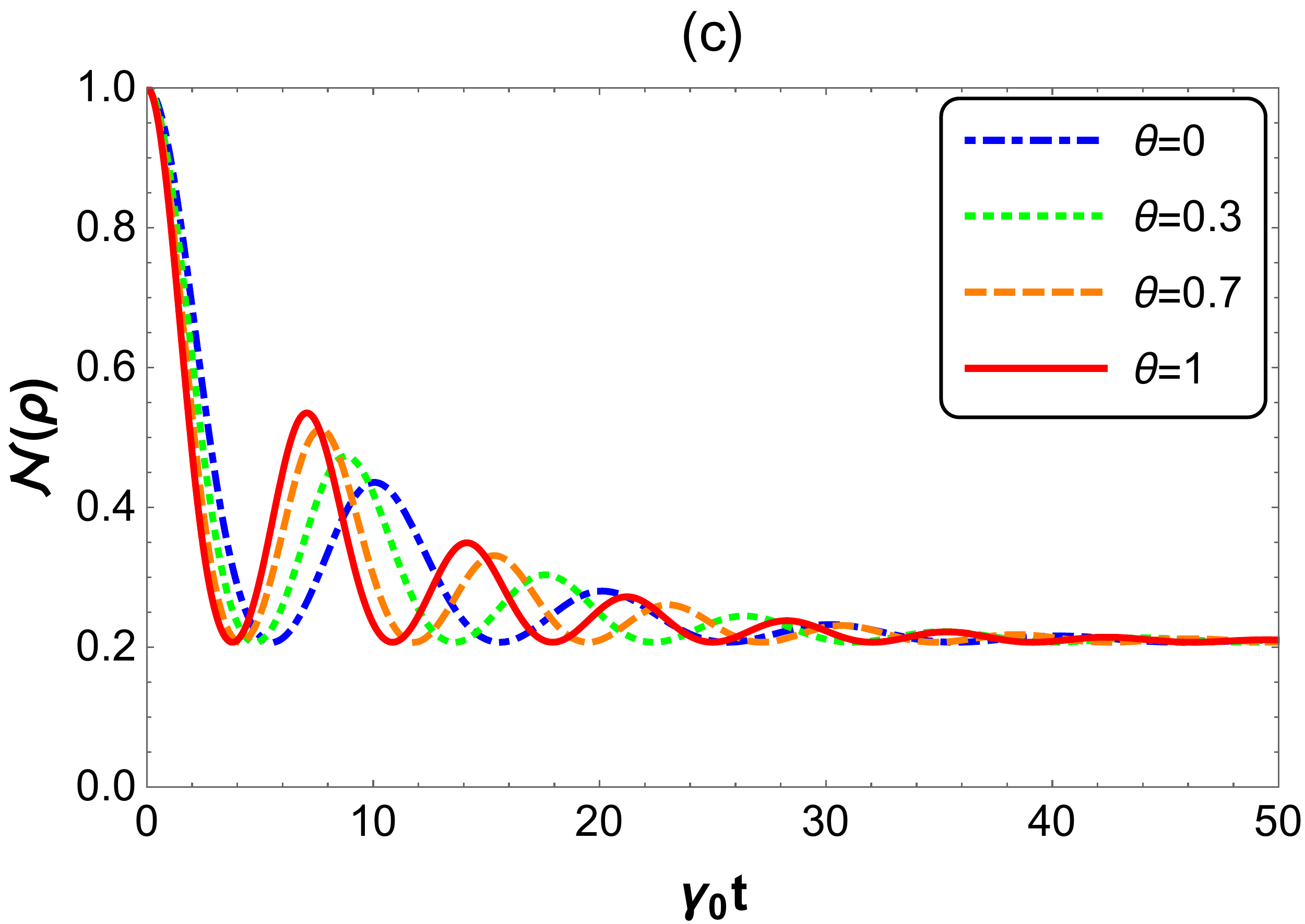}
    \includegraphics[width=4cm,height=3.5cm]{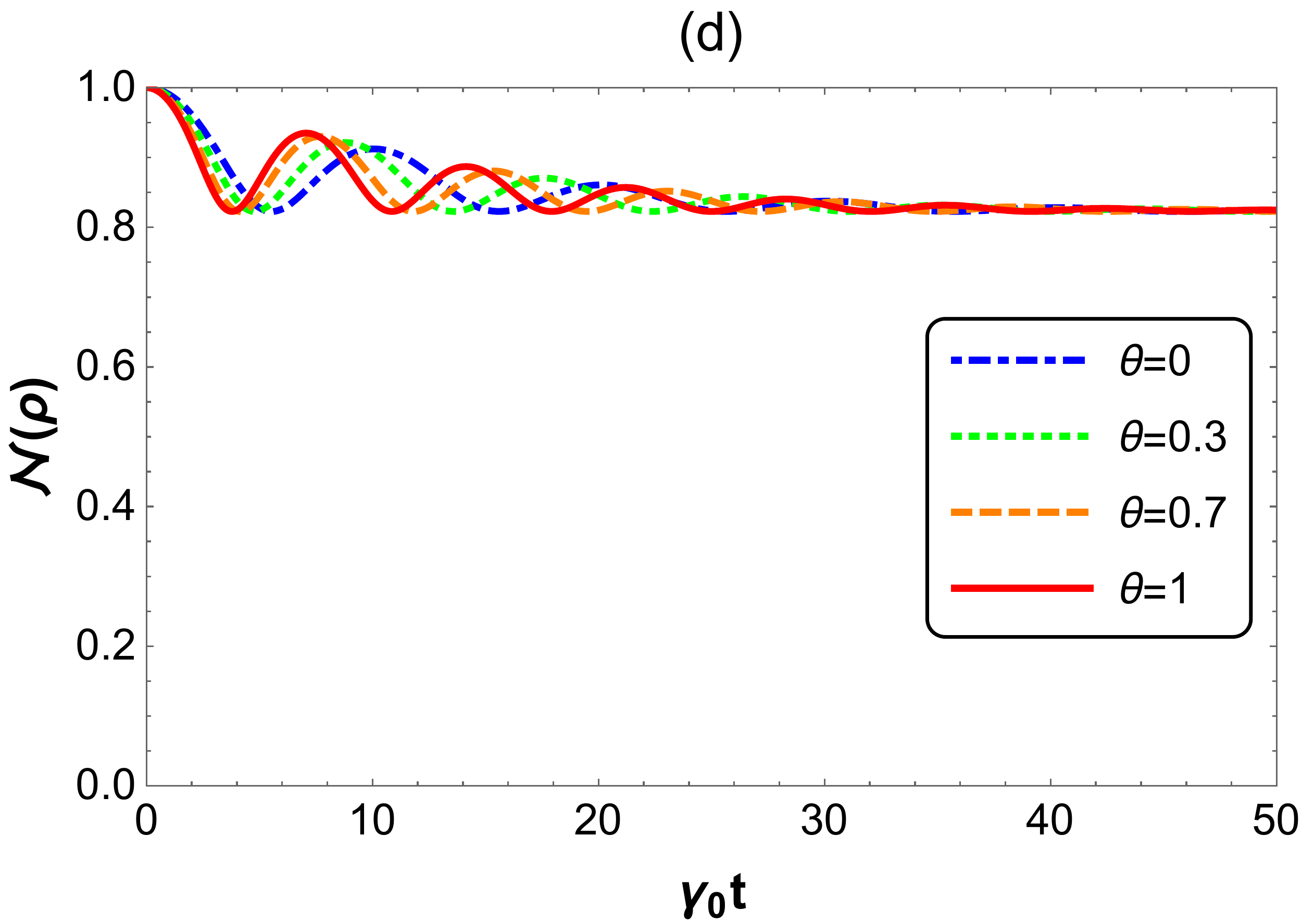}
\caption{(Color online) Negativity in terms of $\gamma_{0}t$ for the initial state $|\psi(0)\rangle=\frac{\sqrt{2}}{2}(|C_{1}A_{2}\rangle+|B_{1}C_{2}\rangle)_{S}\otimes|0\rangle_{R}$ when $\theta =0$ (blue dot dashed line), $\theta =0.3$ (green dotted line), $\theta =0.7$ (orange dashed line) and $\theta =1$ (red solid line). (a),(c) $p_{r}=0$ and (b),(d) $p_{r}=0.9$. (a),(b) $\kappa=10\gamma_{0}$ (with the weak coupling), (c),(d) $\kappa=0.1\gamma_{0}$ (with the strong coupling).   The other parameters are $p=0$, $\gamma_{0}=1$ and $\delta=0$.}
	\label{fig:3}
\end{figure}

In Fig.3, we analyse the effects of SGI parameter and reversing measurement strength on entanglement dynamics in different coupling regimes when the initial state is maximal entangled. Fig.3a indicates that, in the weak coupling regime ($\kappa=10\gamma_{0}$), the negativity quickly decays monotonously from 1 to the steady value 0.2 and its decay rate will become big with $\theta$ increasing if the weak measurement reversal is not performed ($p_{r}=0$). However, when the weak measurement reversal is performed on each atom ($p_{r}=0.9$), the negativity will reduce very slowly from 1 and then tend to 0.83 or not 0.2 even in the weak coupling regime ($\kappa=10\gamma_{0}$ ), as shown in Fig.3b. Comparing Fig.3a and Fig.3b, we see that the weak measurement reversal can effectively protect the entanglement and greatly improve the steady value even in the weak coupling regime. Fig.3c describes the entanglement dynamics when $\kappa=0.1\gamma_{0}$ (in the strong coupling regime) and $p_{r}=0$ (without the weak measurement reversal). We find that the oscillation phenomenon will occur due to the feedback and memory effects of environment, which is the difference from Fig.3a. In addition, a larger SGI parameter corresponds to a bigger amplitude and a smaller period. The steady value in Fig.3c is the same as in Fig.3a. Under the combination of strong coupling ($\kappa=0.1\gamma_{0}$) and weak measurement reversal ($p_{r}=0.9$ ), the entanglement dynamics is shown in Fig.3d. 
On the one hand, the negativity exhibits oscillatory attenuation, which is similar to Fig.3c. On the other hand, the negativity ultimately reaches the steady value 0.82 or not 0.2, which is consistent with Fig.3b. Namely, the smaller SGI parameter, the stronger coupling and the bigger reversing measurement strength can all protect the entanglement, but the weak measurement reversal is more effectively than the cavity-environment coupling and the SGI parameter.
\begin{figure}[tbp]
    \includegraphics[width=4cm,height=3.5cm]{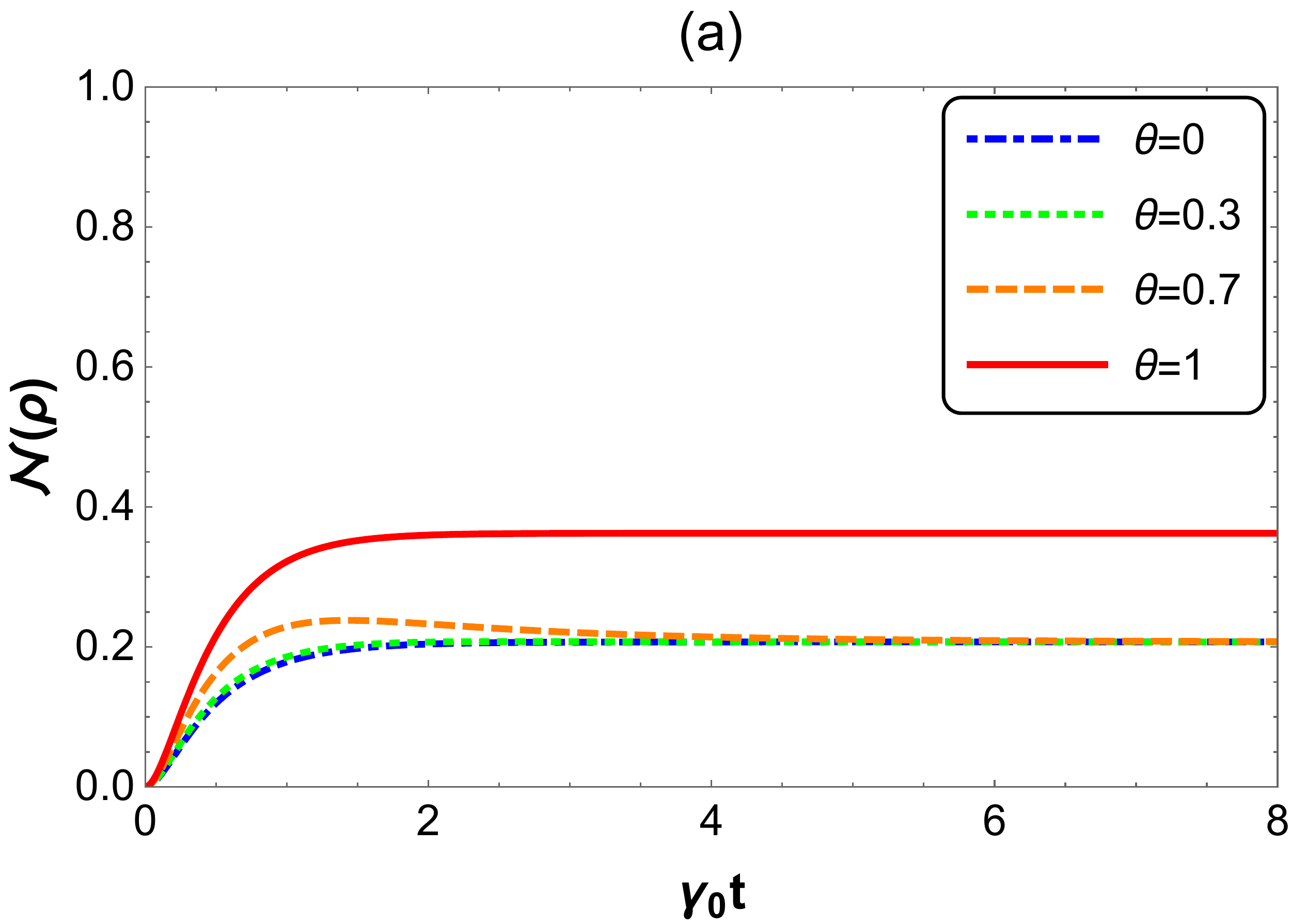}
    \includegraphics[width=4cm,height=3.5cm]{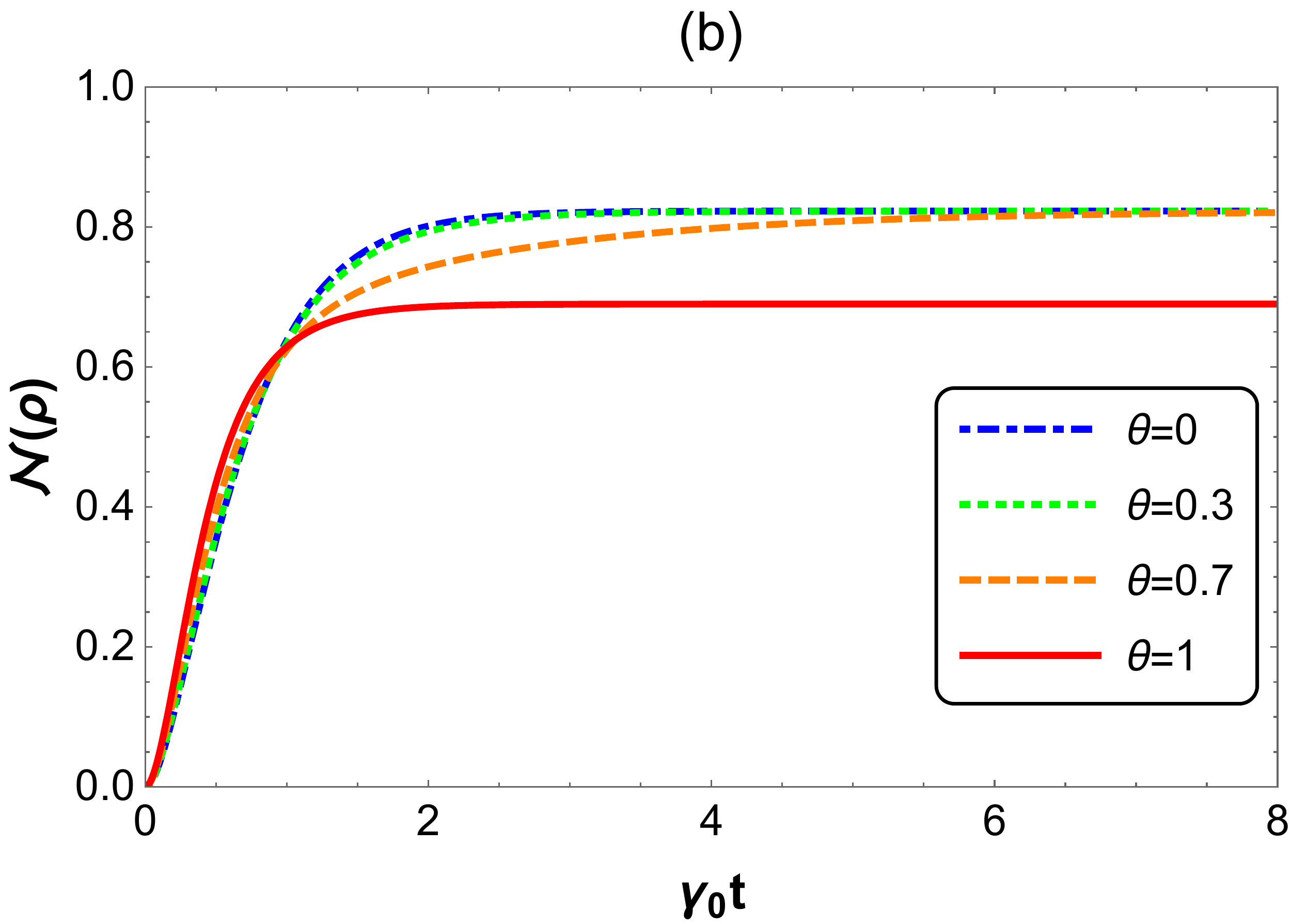}
    \includegraphics[width=4cm,height=3.5cm]{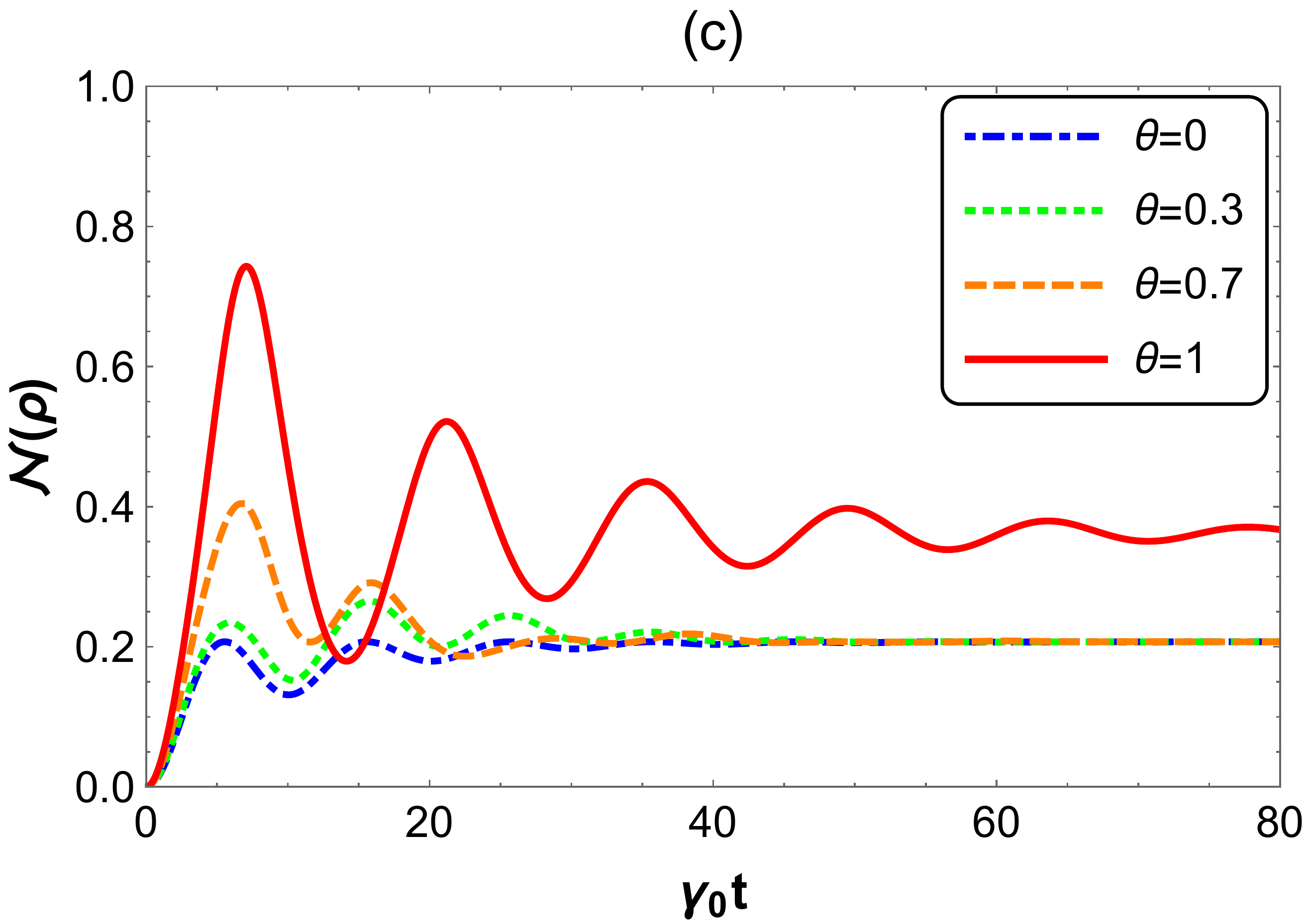}
    \includegraphics[width=4cm,height=3.5cm]{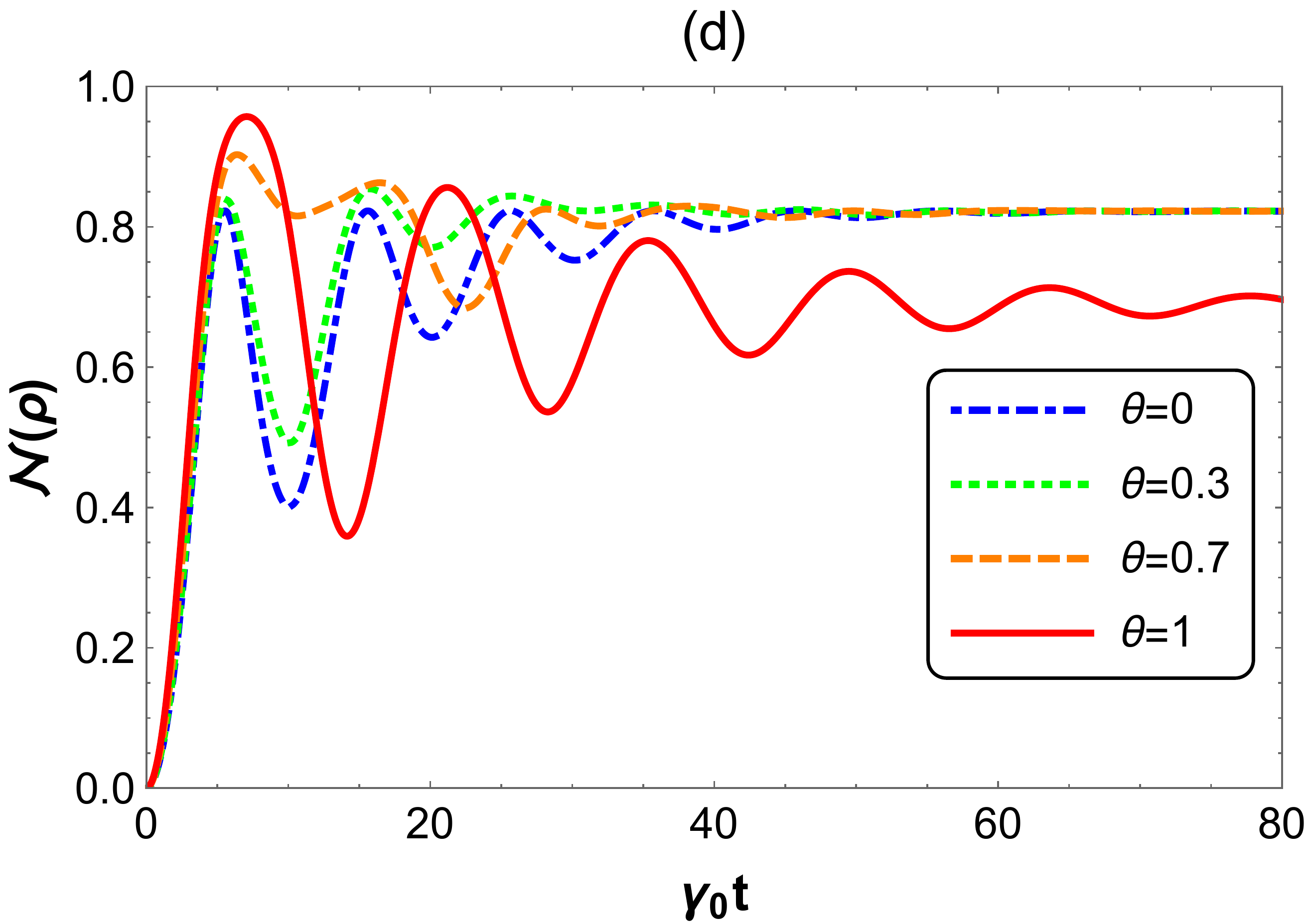}
\caption{(Color online) Negativity in terms of $\gamma_{0}t$ for the initial state $|\psi(0)\rangle=|B_{1}C_{2}\rangle_{S}\otimes|0\rangle_{R}$ when $\theta =0$ (blue dot dashed line), $\theta =0.3$ (green dotted line), $\theta =0.7$ (orange dashed line) and $\theta =1$ (red solid line). (a),(c) $p_{r}=0$ and (b),(d) $p_{r}=0.9$.  (a),(b) $\kappa=10\gamma_{0}$ (with the weak coupling), (c),(d) $\kappa=0.1\gamma_{0}$ (with the strong coupling). . The other parameters are $p=0$, $\gamma_{0}=1$ and $\delta=0$.}
	\label{fig:4}
\end{figure}

Fig.4 gives the effects of SGI parameter, cavity-environment coupling and weak measurement reversal on entanglement dynamics when the two atoms are initial in the product state. Fig.4a shows that, when $\kappa=10\gamma_{0}$ (in the weak coupling regime) and $p_{r}=0$ (without the weak measurement reversal), the entanglement will be generated for all SGI parameters. The larger the SGI parameter is, the faster and more entanglement is generated. In addition, the steady value is 0.20 when $\theta<1$ and is 0.36 when $\theta=1$. In Fig.4b ($p_{r}=0.9$), the entanglement can be generated very quickly and the steady value is 0.82 when $\theta< 1$ and is 0.69 when $\theta=1$, which is obviously greater than that in Fig.4a. Namely, the weak measurement reversal can generate more entanglement in the product state. Fig.4c shows that, in the strong coupling regime ($\kappa=0.1\gamma_{0}$ ), the entanglement will be generated quickly and reach a peak value and then oscillate to a steady value due to the feedback and memory effects of environment, which is the main difference from Fig.4a. The peak and period of entanglement will become larger with $\theta$ increasing, for example, the maximal peak is 0.74 when $\theta=1$ but it is 0.21 when $\theta=0$. In addition, the steady value is 0.20 when $\theta<1$ and is 0.36 when $\theta=1$, which is the same as Fig.4a. Under the combination of strong coupling  ($\kappa=0.1\gamma_{0}$ ) and weak measurement reversal ($p_{r}=0.9$),  the entanglement dynamics is shown in Fig.4d. The entanglement will be generated very quickly and then oscillate to a steady value. The entanglement peak in Fig.4d are obviously greater than those in Fig.4c and the steady values in Fig.4d are the same as in Fig.4b. Therefore, the smaller SGI parameter, the stronger coupling and the bigger reversing measurement strength can all generate the entanglement in the product state, and the weak measurement reversal can generate more entanglement than the coupling and the SGI parameter.
\begin{figure}[tbp]
    \includegraphics[width=4cm,height=3.5cm]{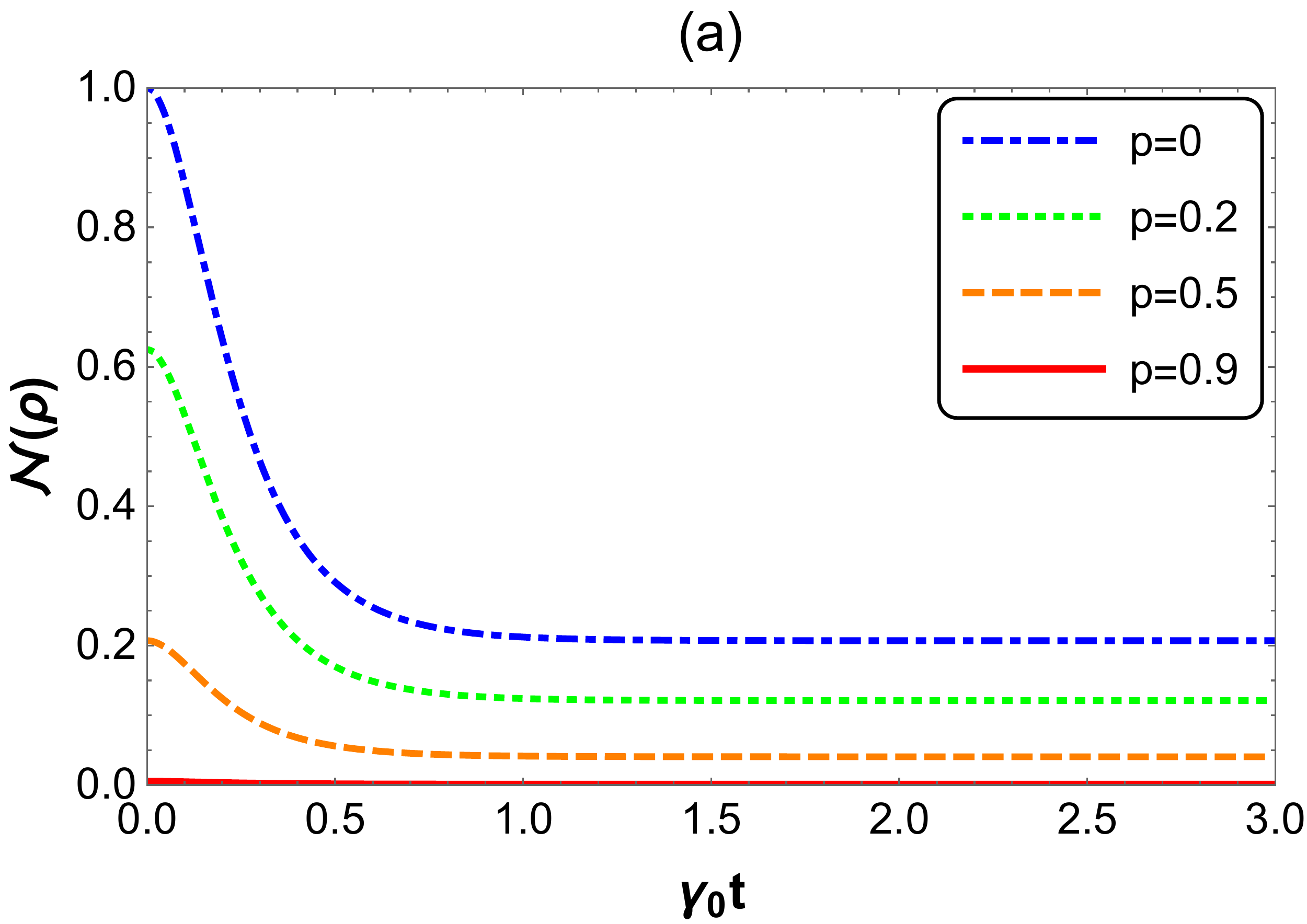}
    \includegraphics[width=4cm,height=3.5cm]{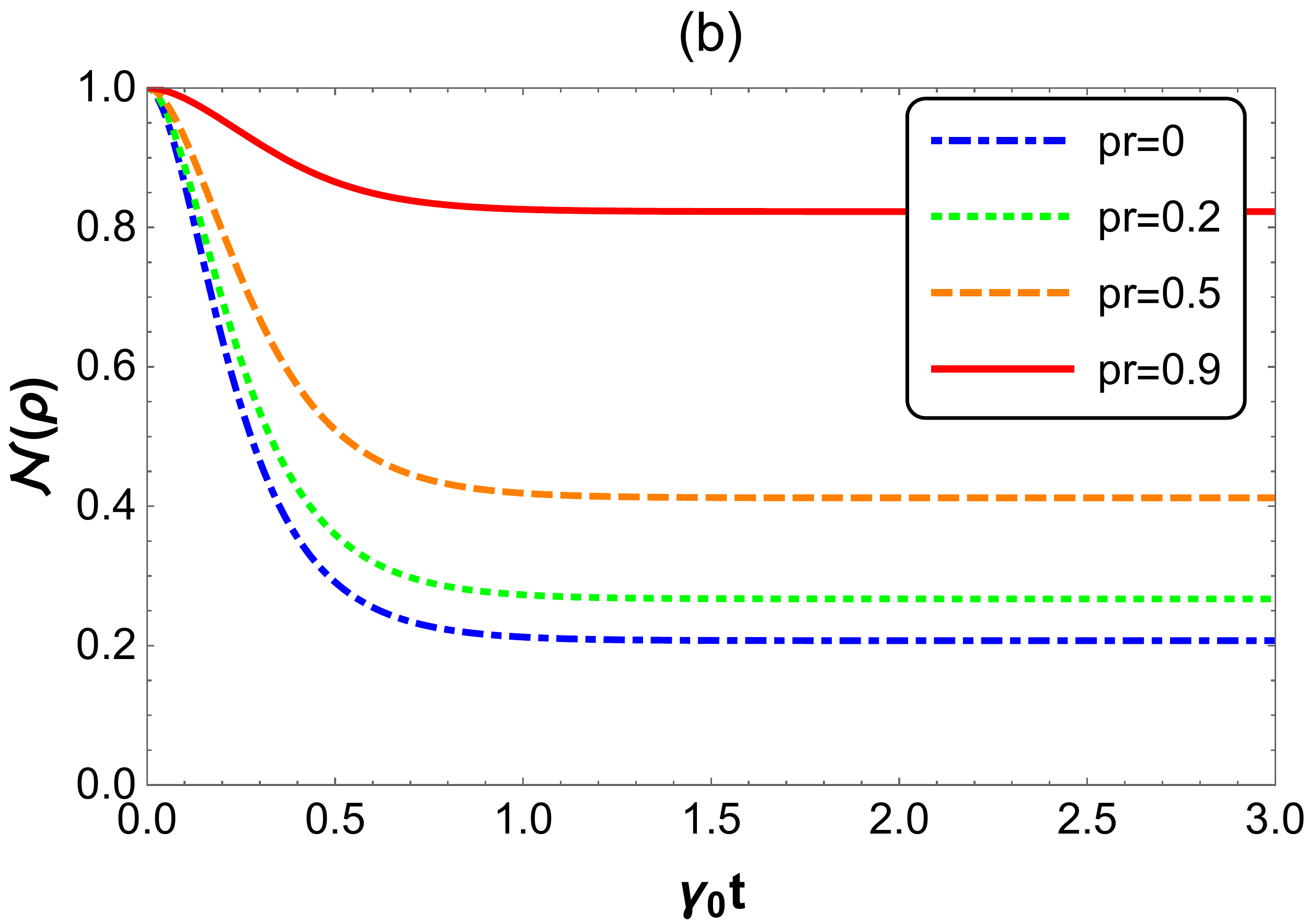}
    \includegraphics[width=4cm,height=3.5cm]{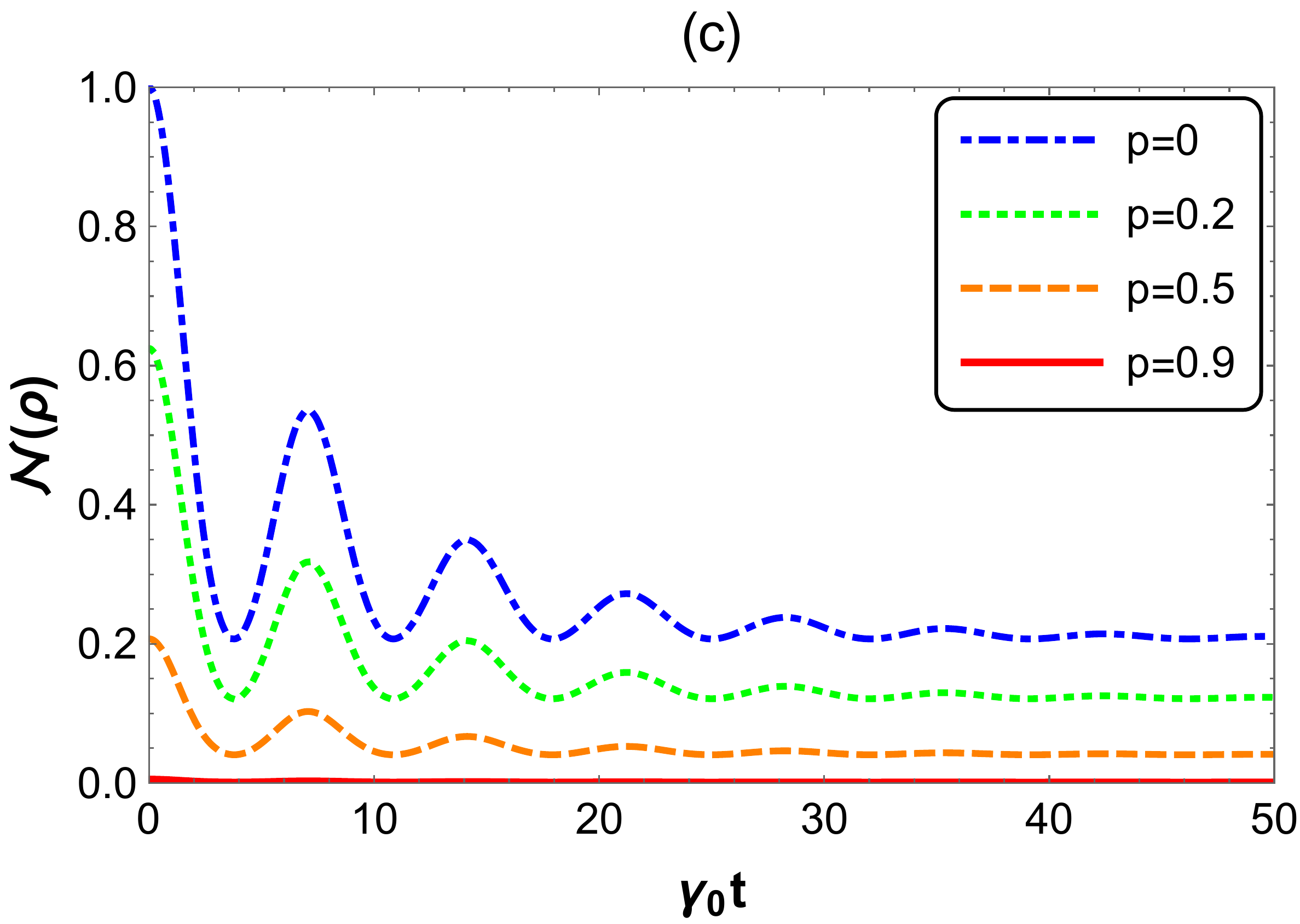}
    \includegraphics[width=4cm,height=3.5cm]{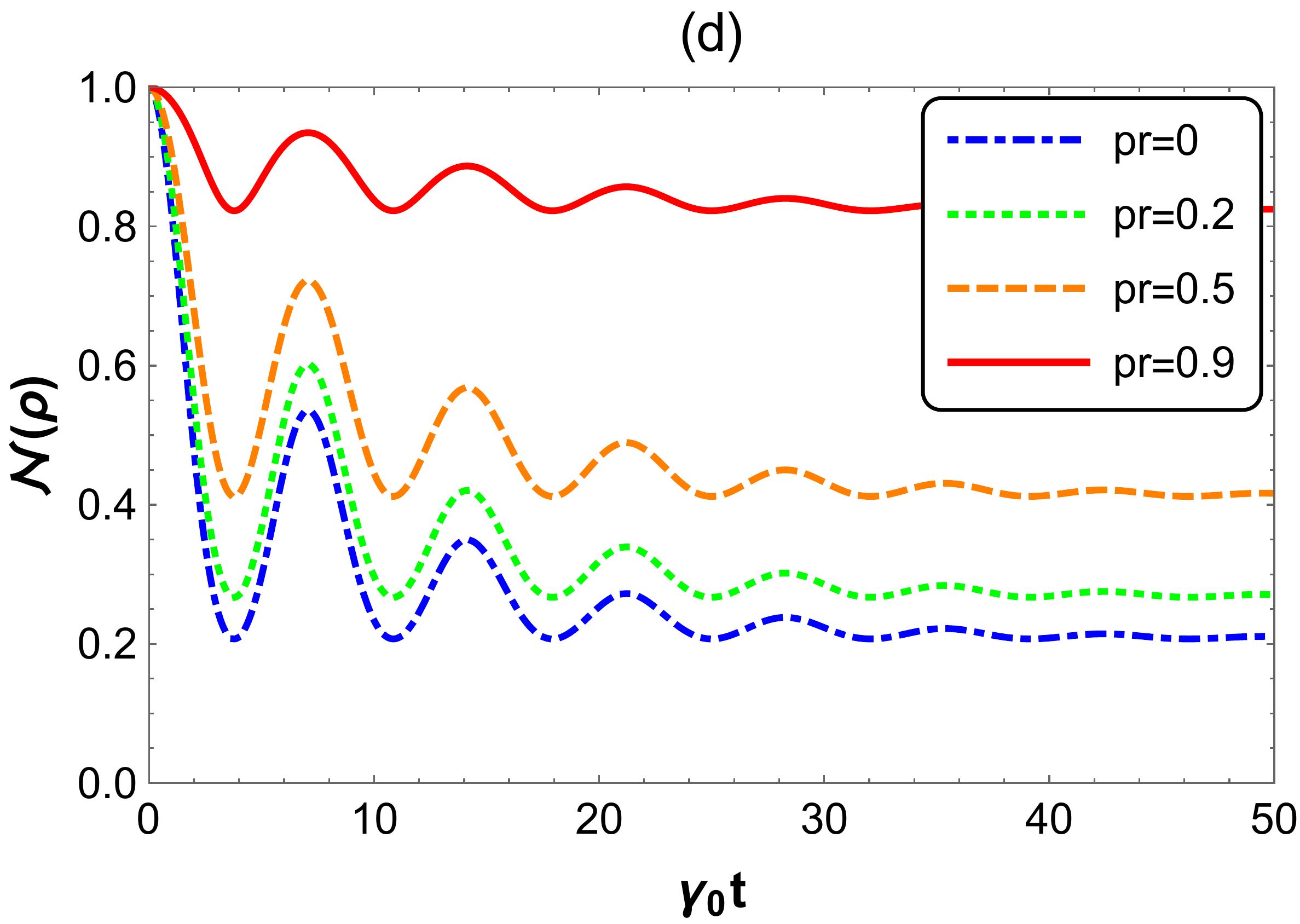}
\caption{(Color online)  Negativity as a function of  $\gamma_{0}t$ for the initial state $|\psi(0)\rangle=\frac{\sqrt{2}}{2}(|C_{1}A_{2}\rangle+|B_{1}C_{2}\rangle)_{S}\otimes|0\rangle_{R}$. (a),(c) $p=0$ (blue dot dashed line), $p=0.2$ (green dotted line), $p=0.5$ (orange dashed line) and $p=0.9$ (red solid line), $p_{r}=0$. (b),(d) $p_{r}=0$ (blue dot dashed line), $p_{r}=0.2$ (green dotted line), $p_{r}=0.5$ (orange dashed line) and $p_{r}=0.9$ (red solid line), $p=0$. (a),(b) $\kappa=10\gamma_{0}$ (with the weak coupling), (c),(d) $\kappa=0.1\gamma_{0}$ (with the strong coupling). The other parameters are $\theta =1$, $\gamma_{0}=1$ and $\delta=0$.}
	\label{fig:5}
\end{figure}

The dynamical behavior of entanglement in terms of dimensionless time $\gamma_{0}t$ has been shown in Fig.5 for different weak and reversing measurement strengths when the initial state is maximal entangled. Fig.5a indicates that, in the weak coupling regime, the initial and steady values of negativity all reduce with $p$ increasing when $p_{r}=0$. Specially, the negativity is close to zero when $p=0.9$. But, the negativity always decays monotonously from 1 for different $p_{r}$. The larger $p_{r}$ is, the slower the negativity decays and the larger the steady value is (see Fig 5b). Fig.5c shows that, in the strong coupling regime, the negativity will oscillate and the amplitude becomes small with $p$ increasing, which is different from Fig.5a. The initial and steady values in Fig.5c are the same as Fig.5a for different $p$. Comparing Fig.5d with Fig.5b, their difference is what the former will exhibit oscillation phenomena but the latter decreases  monotonously, and the steady values are the same for both. Hence, for the maximal entangled state, whether in weak or strong coupling regimes, the entanglement will be destroyed by the weak measurement but protected by the weak measurement reversal. 
\begin{figure}[tbp]
    \includegraphics[width=4cm,height=3.5cm]{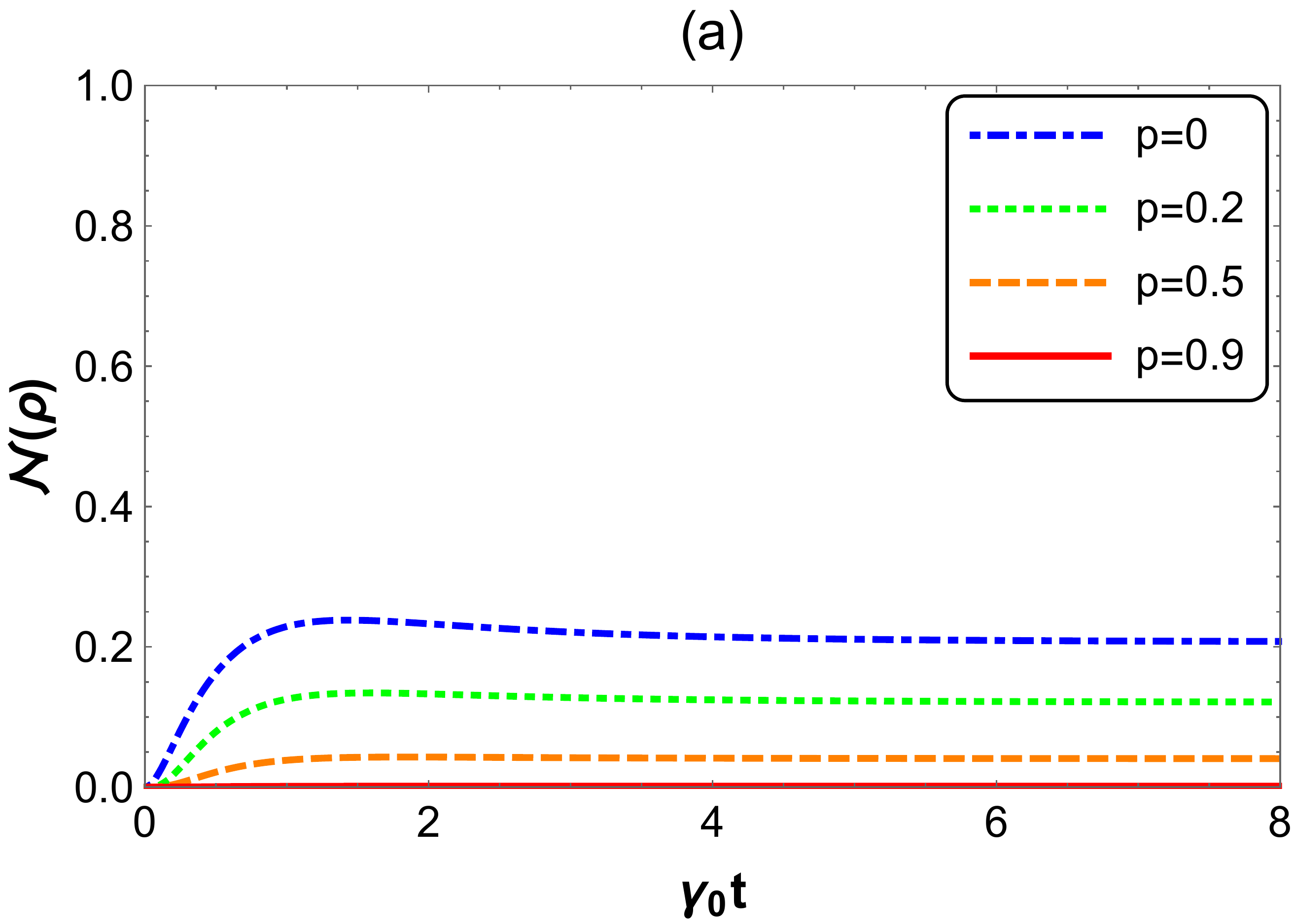}
    \includegraphics[width=4cm,height=3.5cm]{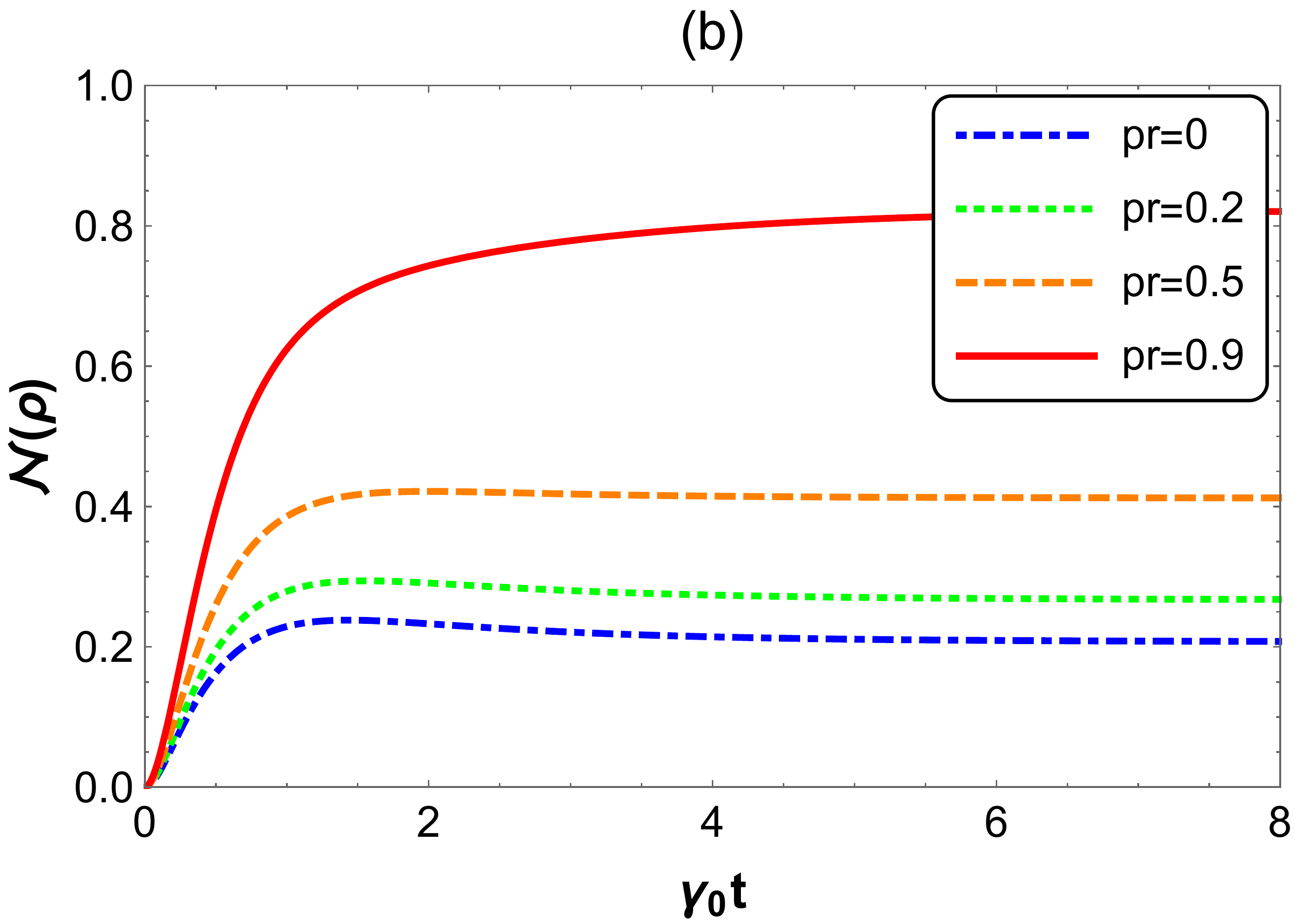}
    \includegraphics[width=4cm,height=3.5cm]{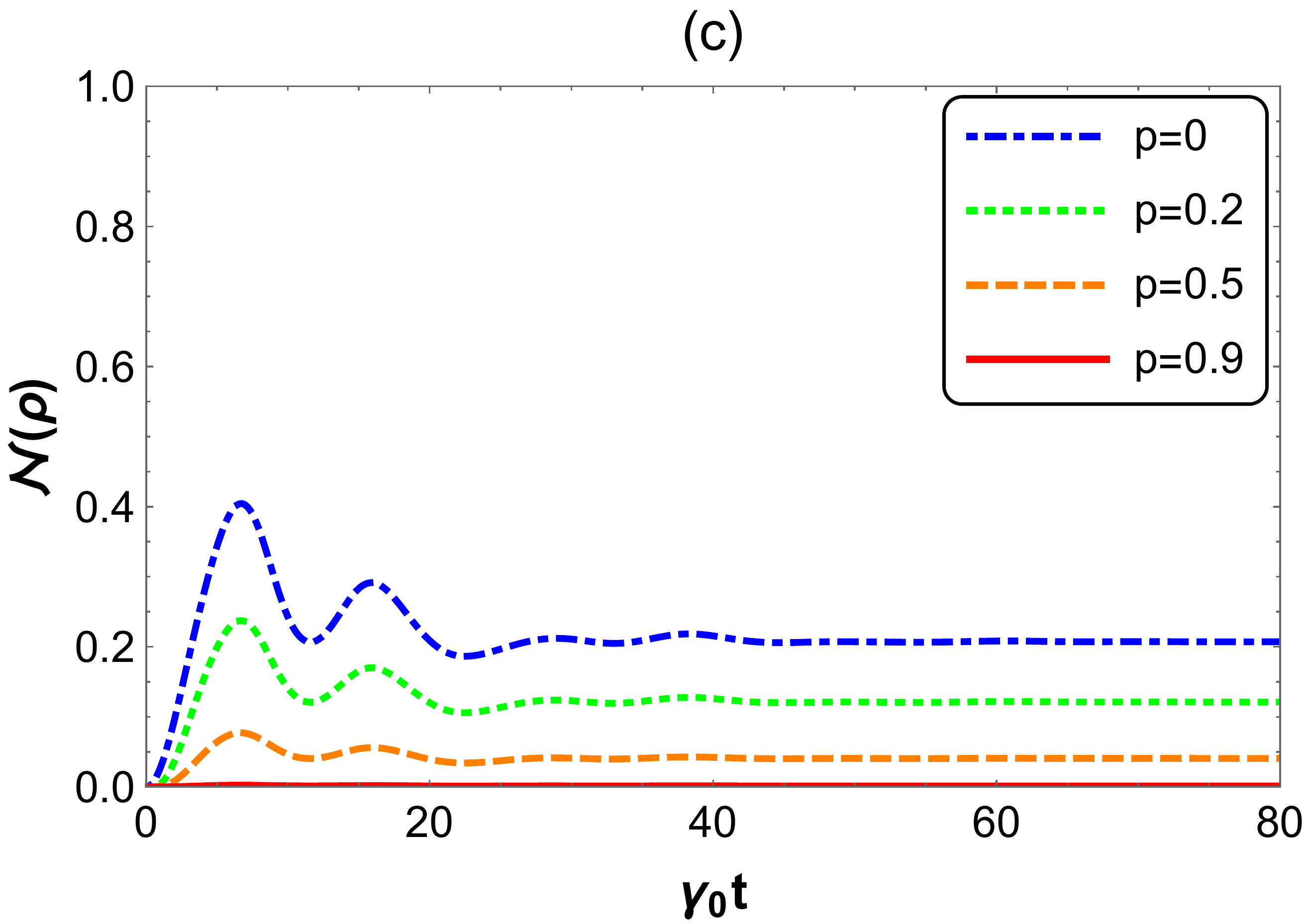}
    \includegraphics[width=4cm,height=3.5cm]{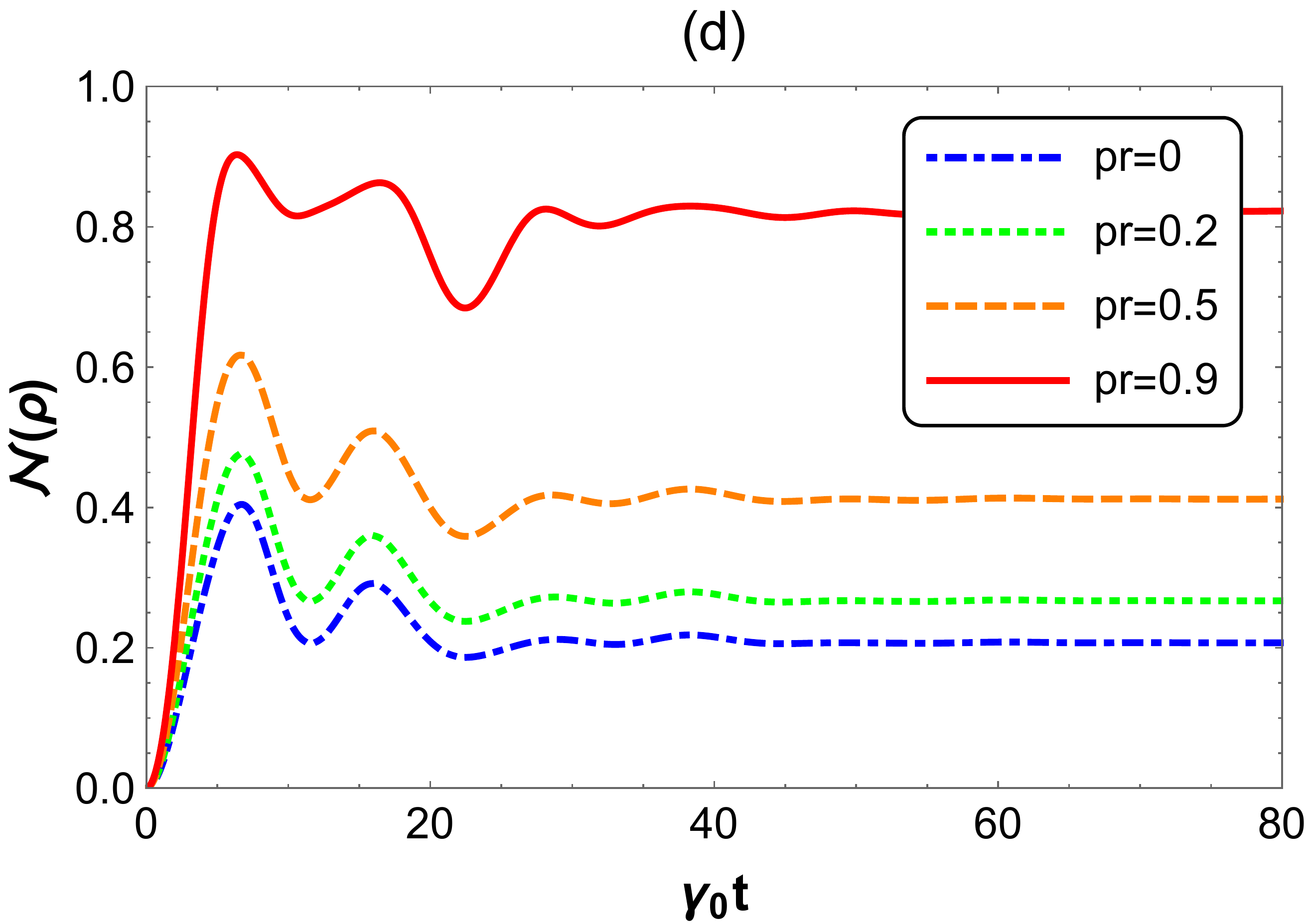}
\caption{(Color online) Negativity as a function of  $\gamma_{0}t$ for the initial state $|\psi(0)\rangle=|B_{1}C_{2}\rangle_{S}\otimes|0\rangle_{R}$. (a),(c) $p=0$ (blue dot dashed line), $p=0.2$ (green dotted line), $p=0.5$ (orange dashed line) and $p=0.9$ (red solid line), $p_{r}=0$. (b),(d) $p_{r}=0$ (blue dot dashed line), $p_{r}=0.2$ (green dotted line), $p_{r}=0.5$ (orange dashed, line) and $p_{r}=0.9$ (red solid line), $p=0$. (a),(b) $\kappa=10\gamma_{0}$ (with the weak coupling), (c),(d) $\kappa=0.1\gamma_{0}$ (with the strong coupling). The other parameters are $\theta =0.7$, $\gamma_{0}=1$ and $\delta=0$.}
	\label{fig:6}
\end{figure}

Fig.6 exhibits the effects of weak and reversing measurement strengths on entanglement dynamics in different coupling regimes when the two atoms are in the product state. From the blue dot-dashed line in Fig.6a, we find that, in the weak coupling regime, the negativity will grow to 0.2 from zero when $p=0$ and the less entanglement is generated as $p$ increases, for instance, $N(\rho)\to 0$ when $p=0.9$. Fig.6b shows that, the negativity will also grow to 0.2 from zero when $p_{r}=0$ but the more entanglement is generated as $p_{r}$ enlarges, for instance, $N(\rho)\to 0.82$ when $p_{r}=0.9$. In Fig.6c, the negativity will grow and then oscillate to a steady value, and a bigger $p$ corresponds to a smaller peak, but Fig.6c has the same steady value as Fig.6a for each $p$. Under the combination of strong coupling and weak measurement reversal, the entanglement dynamics is described in Fig.6d. The entanglement grows to a peak from 0 and then oscillates to a steady value, and a bigger $p_{r}$ corresponds to a greater peak, but Fig.6d has the same steady value as Fig.6b for each $p_{r}$. Thus, for the product state, whether in the weak or strong coupling regimes, the generation of entanglement can be suppressed by the weak measurement but be enhanced very effectively by the weak measurement reversal. 

\subsection{ Entanglement dynamics with detuning}
Nextly, we consider the entanglement dynamics with detuning ($\delta \ne 0$).
 \begin{figure}[tbp]
    \includegraphics[width=4cm,height=3.5cm]{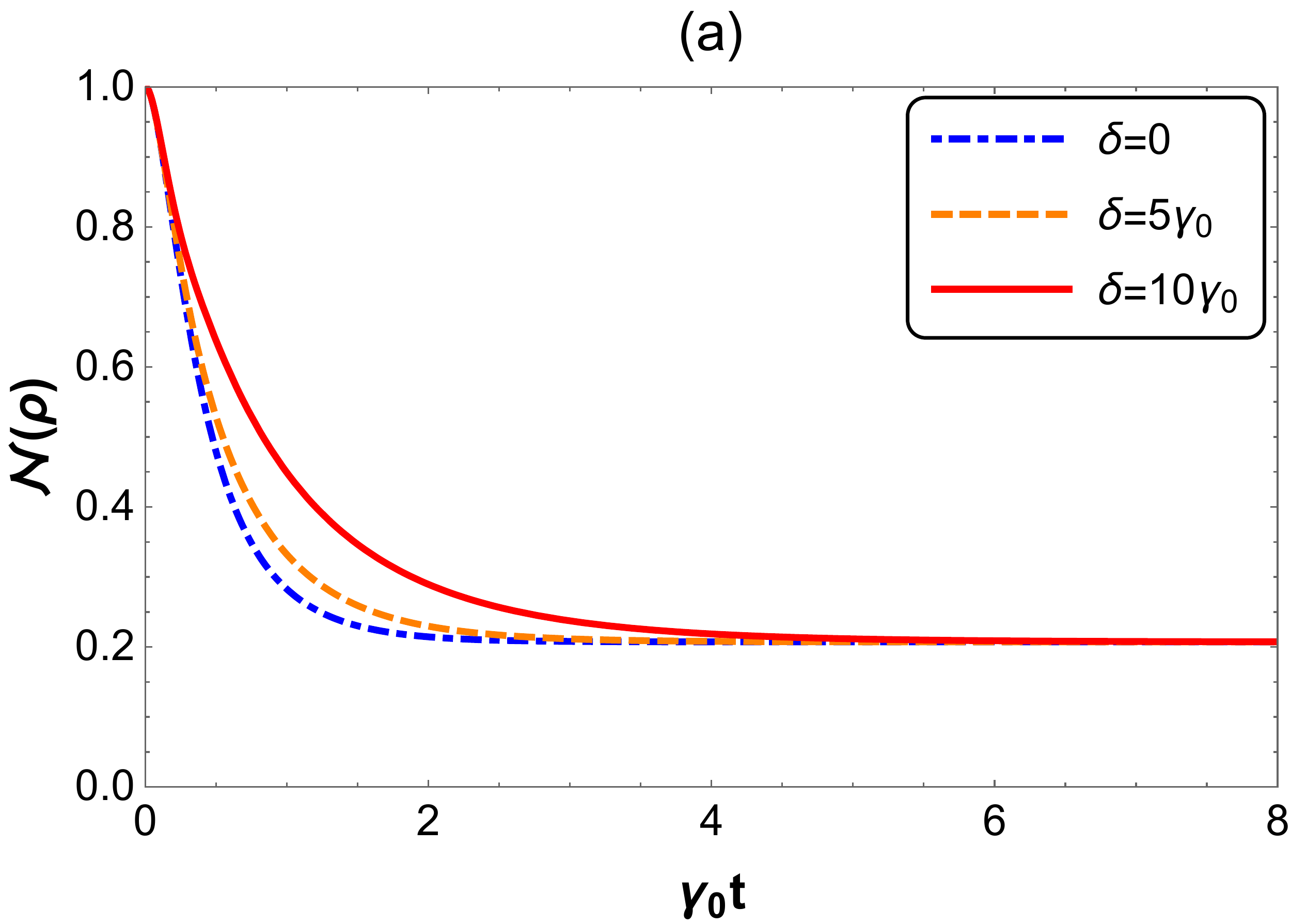}
    \includegraphics[width=4cm,height=3.5cm]{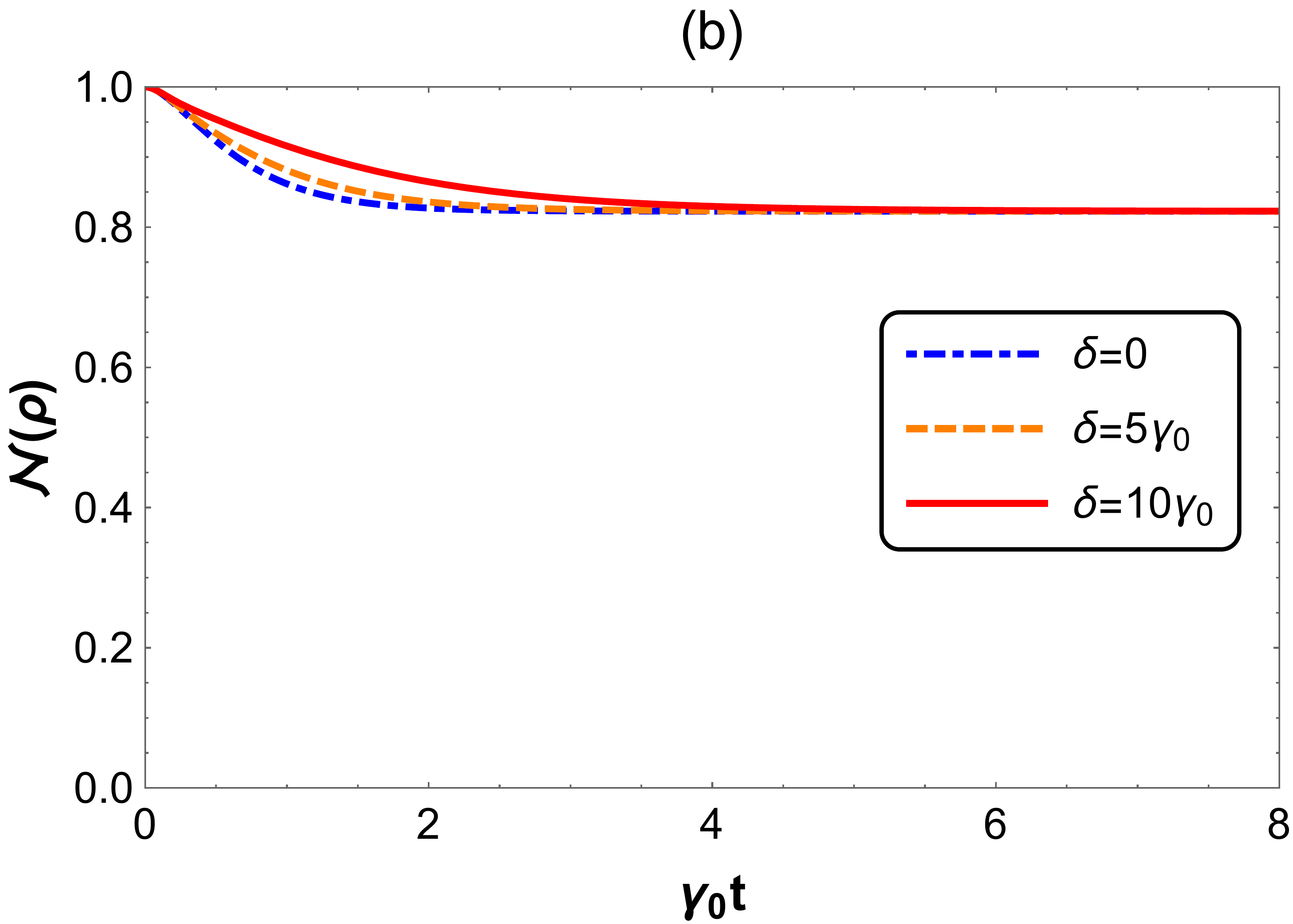}
    \includegraphics[width=4cm,height=3.5cm]{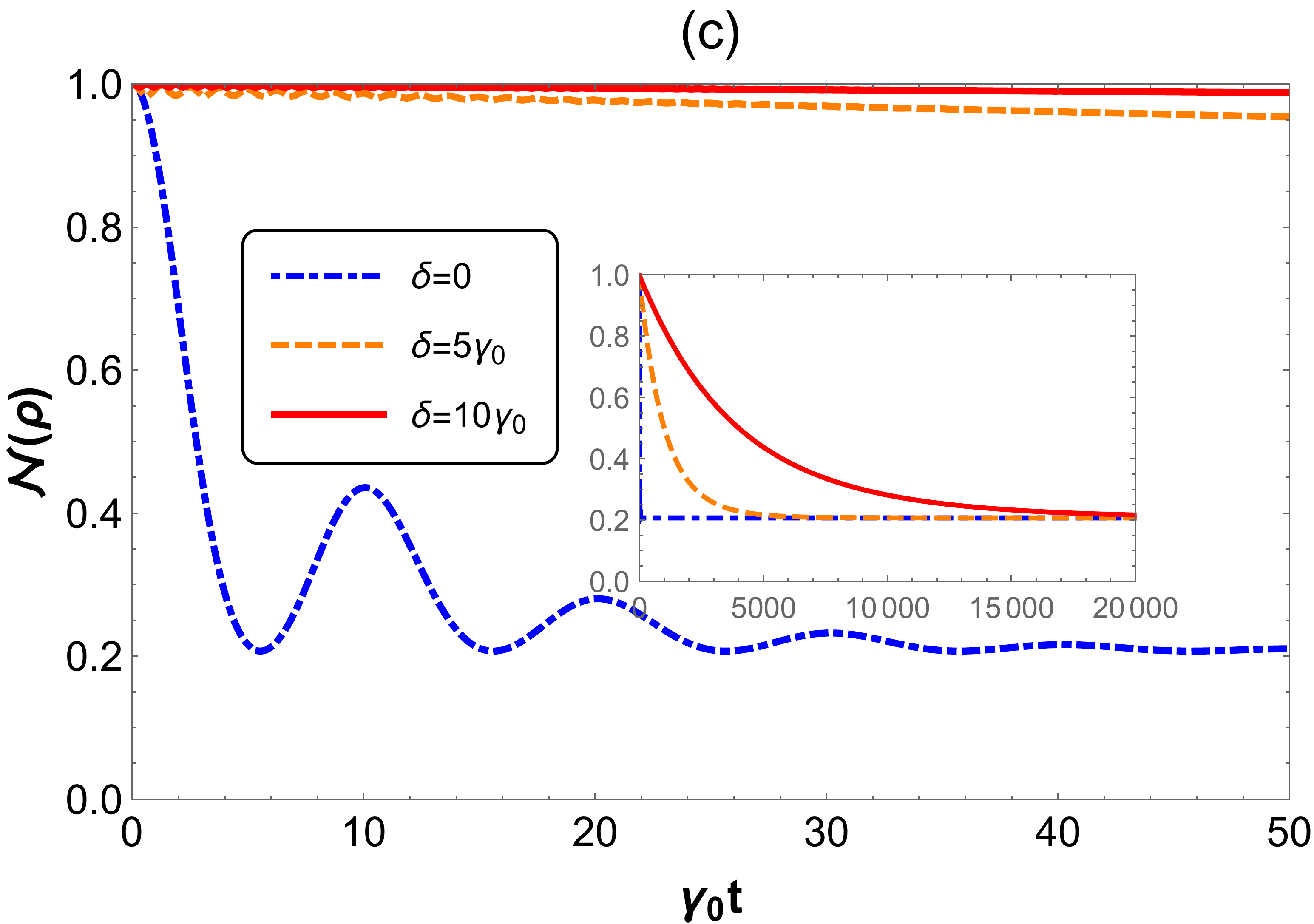}
    \includegraphics[width=4cm,height=3.5cm]{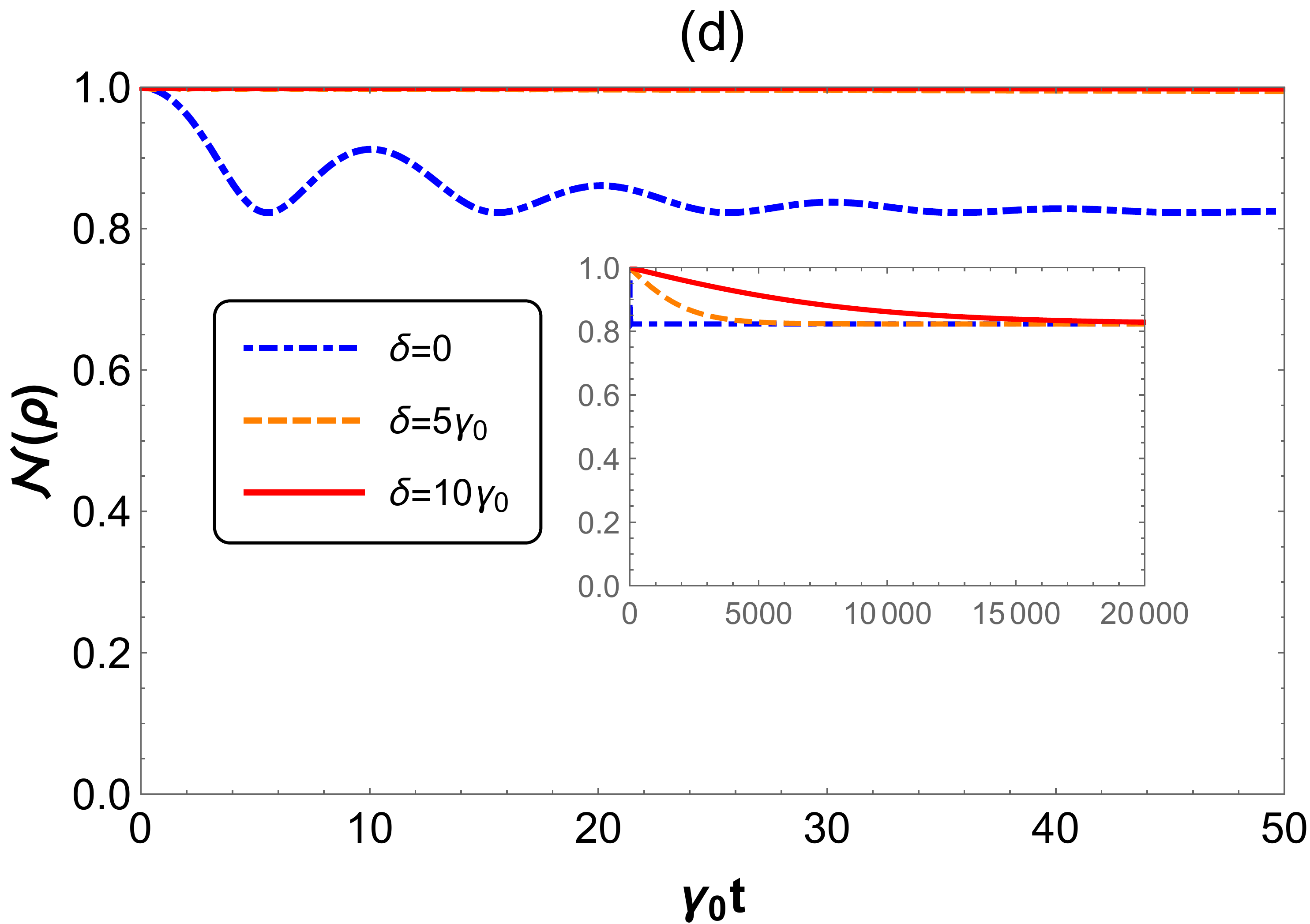}
    
	\caption{(Color online) Negativity as a function of time $\gamma_{0}t$ for the initial state $|\psi(0)\rangle=\frac{\sqrt{2}}{2}(|C_{1}A_{2}\rangle+|B_{1}C_{2}\rangle)_{S}\otimes|0\rangle_{R}$ when $\delta=0$ (blue dot dashed line), $\delta=5\gamma_{0}$ (orange dashed line) and $\delta=10\gamma_{0}$ (red solid line). (a),(c) $p_{r}=0$ and (b),(d) $p_{r}=0.9$. (a),(b) $\kappa=10\gamma_{0}$ (with the weak coupling), (c),(d) $\kappa=0.1\gamma_{0}$ (with the strong coupling). The other parameters are $p=0$, $\theta =0$ and $\gamma_{0}=1$.}
	\label{fig:7}
\end{figure}

Fig.7 gives the influence of detuning on evolution dynamics of the maximal entangled state in different coupling regimes. Fig.7a ($\kappa=10\gamma_{0})$ shows that the negativity will reduce monotonically to 0.2 and its decay rate will become small with $\delta$ increasing when $p_{r}=0$. However, the negativity will reduce to 0.82 or not 0.2 when $p_{r}=0.9$ and $\kappa=10\gamma_{0}$ (see Fig 7b).  From Fig.7c ($\kappa=0.1\gamma_{0}$), we see that, the negativity will quickly reduce and then oscillate to 0.2 when $\delta=0$ (see the blue dotted line in Fig.7c). When $\delta=5\gamma_0$, the negativity will reduce very slowly (see the orange dashed line in Fig.7c). In particular, when $\delta=10\gamma_0$, the negativity will always approaches 1.0 with time increasing (see the orange dashed line in Fig.7c). Only after a very long enough time, the negativity will tend to the steady value 0.2 (see the subgraph inserted in Fig. 7c). Under the combination of strong coupling and weak measurement reversal, the maximal entangled state can be very well protected by the detuning, as shown the orange and red lines in Fig.7d. After a very long enough time, the negativity will tend to the steady value 0.82 (see the subgraph inserted in Fig.7d). Thus, in the weak coupling regime, the protection of entanglement mainly depends on the weak measurement reversal. But the detuning protects the entanglement more effectively than the weak measurement reversal in the strong coupling regime.

\begin{figure}[tbp]
    \includegraphics[width=4cm,height=3.5cm]{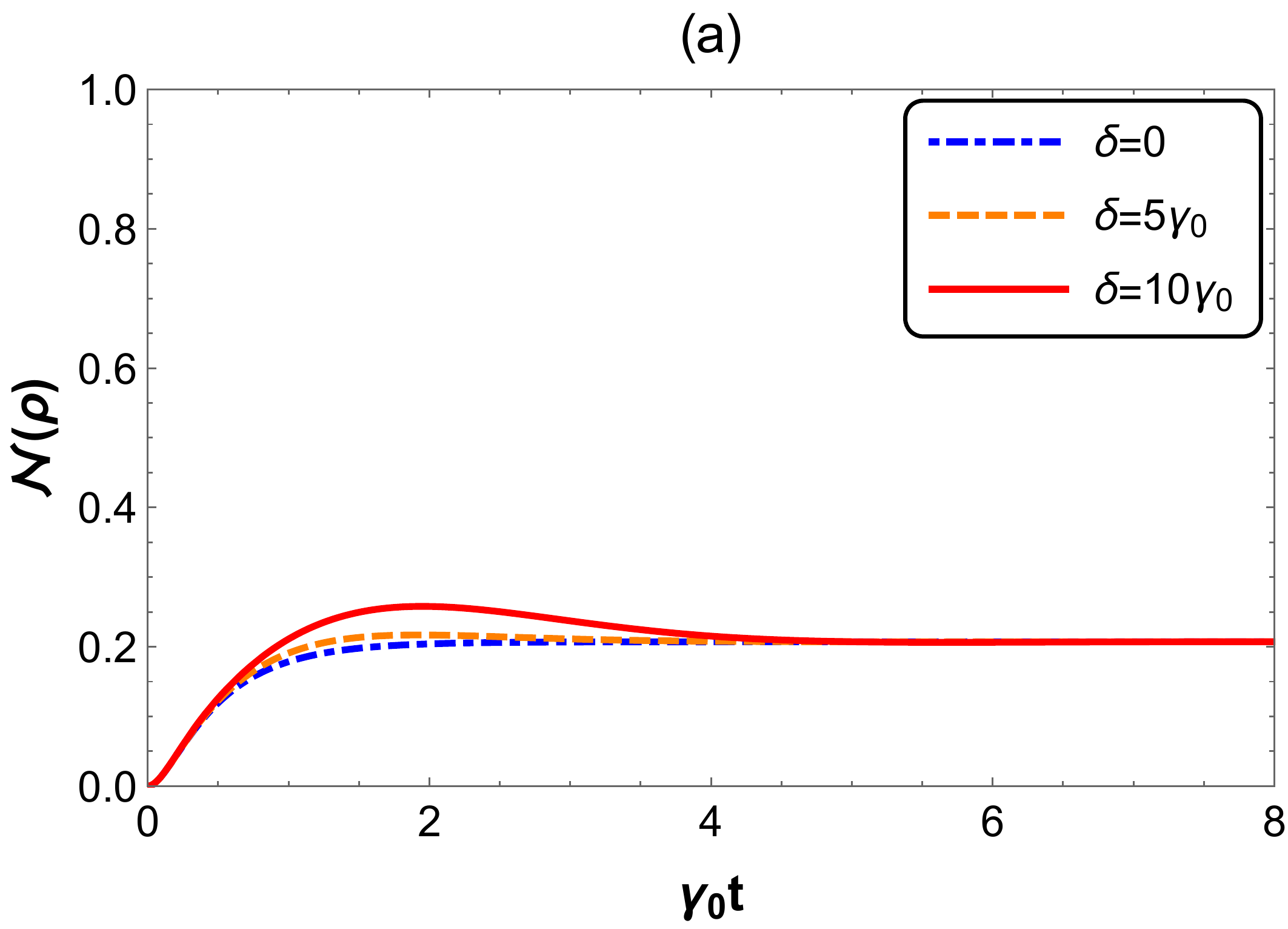}
    \includegraphics[width=4cm,height=3.5cm]{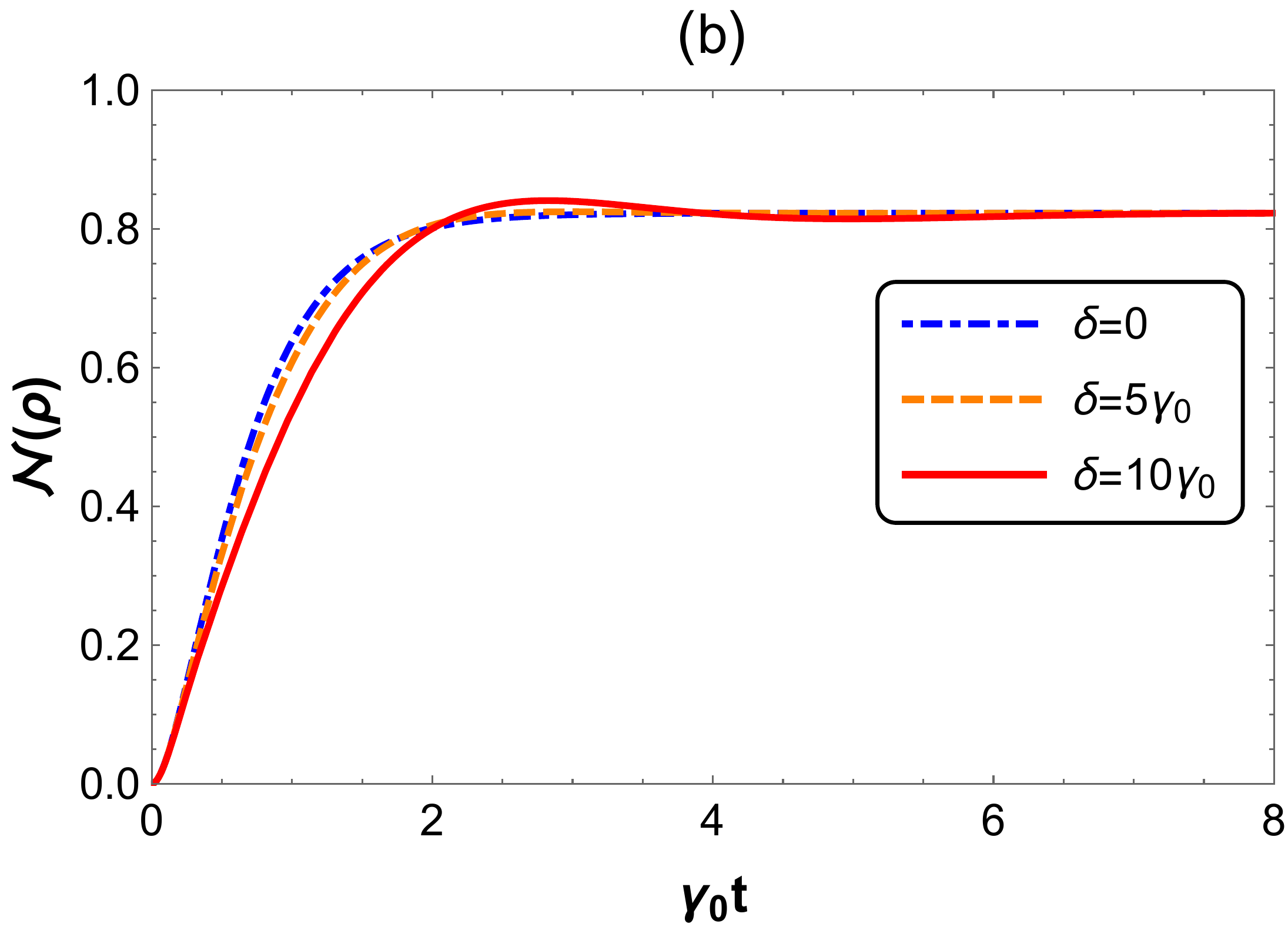}
    \includegraphics[width=4cm,height=3.5cm]{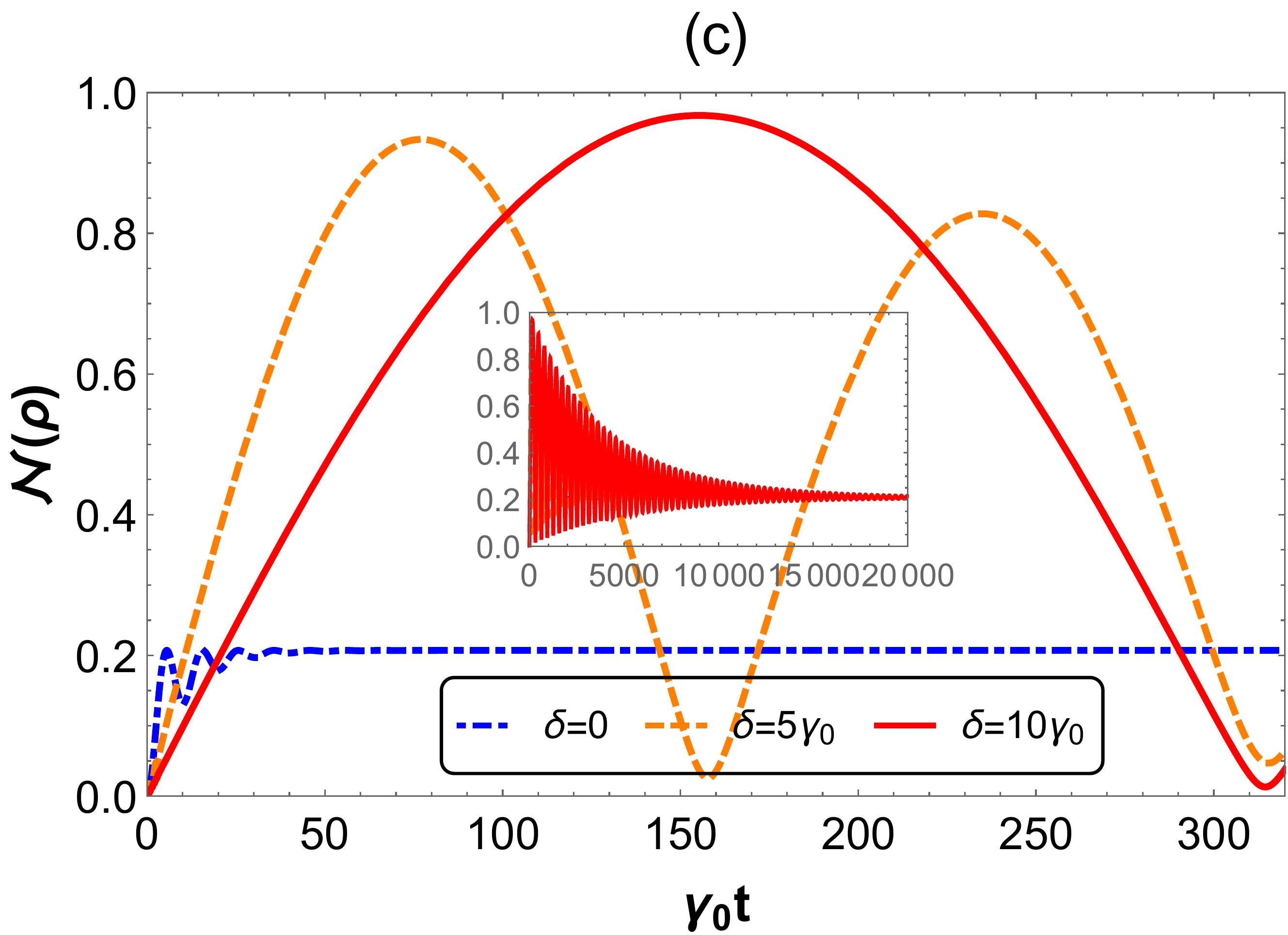}
    \includegraphics[width=4cm,height=3.5cm]{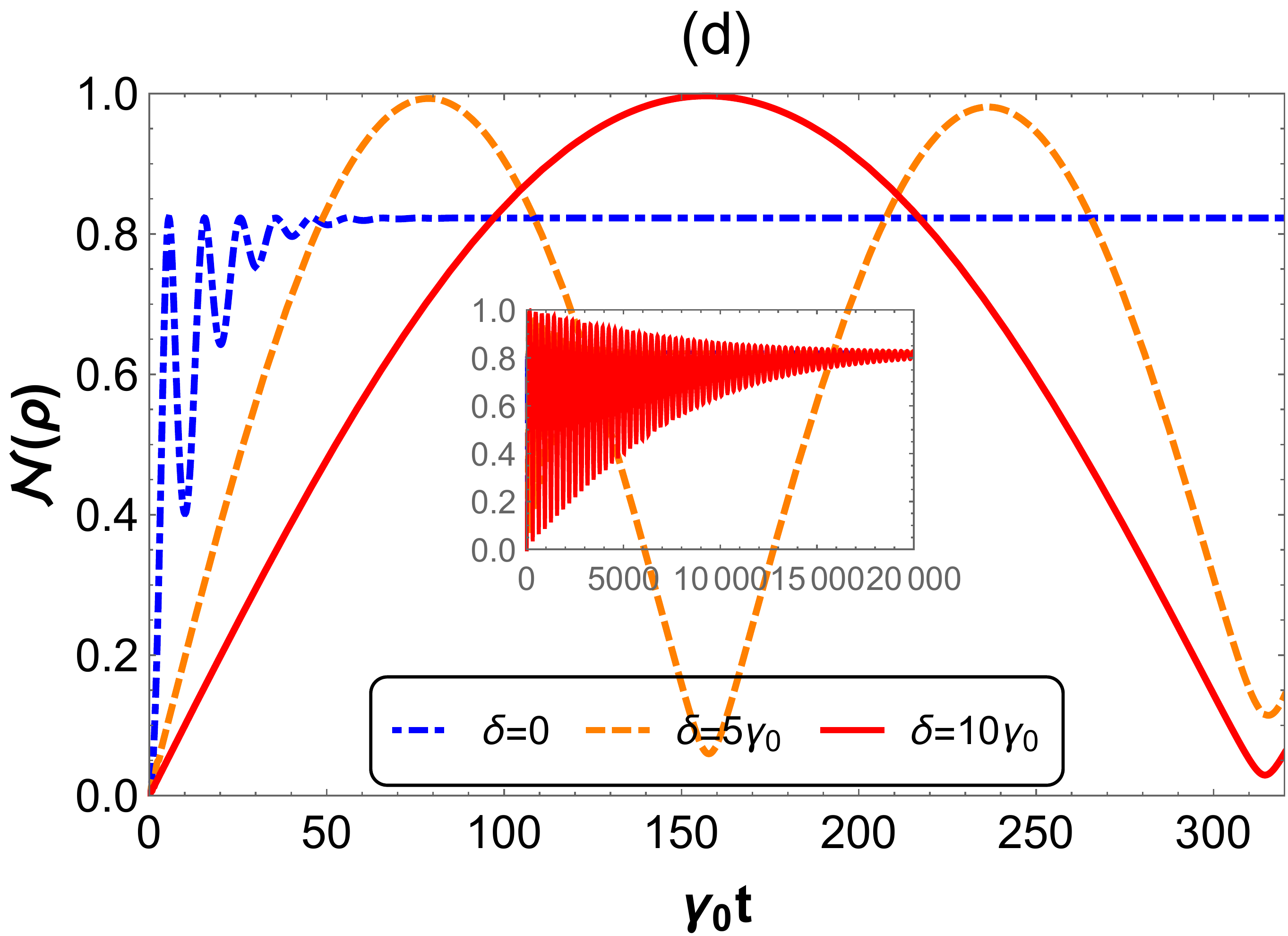}
	\caption{(Color online) Negativity as a function of time $\gamma_{0}t$ for the initial state $|\psi(0)\rangle=|B_{1}C_{2}\rangle_{S}\otimes|0\rangle_{R}$ when $\delta=0$ (blue dot dashed line), $\delta=5\gamma_{0}$ (orange dashed line) and $\delta=10\gamma_{0}$ (red solid line). (a),(c) $p_{r}=0$ and (b),(d) $p_{r}=0.9$. (a),(b) $\kappa=10\gamma_{0}$ (with the weak coupling), (c),(d) $\kappa=0.1\gamma_{0}$ (with the strong coupling). The other parameters are $p=0$, $\theta =0$ and $\gamma_{0}=1$.}
	\label{fig:8}
\end{figure}

In order to demonstrate the influence of detuning on the product state, we draw the entanglement dynamics curves in different coupling regimes (see Fig.8). From Fig.8a, we can see that the negativity will grow to 0.2 from zero when $\kappa=10\gamma_0$ (in the weak coupling regime) and $p_{r}=0$ (without the weak measurement reversal). If $\kappa=10\gamma_0$ and $p_{r}=0.9$, the negativity will grow to 0.82 from zero, as shown in Fig.8b. Fig.8c ($p_{r}=0$) shows that, the negativity will grow to 0.2 from zero when $\delta=0 $ but it will grow to 0.92 from zero and then oscillate to 0.2 when $\delta =5\gamma_0$. The bigger the detuning is, the greater the entanglement peak and the oscillation period are. In Fig.8d ($p_{r}=0.9$), we find that the negativity will grow to 0.82 from zero when $\delta=0 $ but it will grow to 1.0 from zero and then oscillate to 0.82 when $\delta=10\gamma_0$. In the same way, a bigger detuning corresponds to a greater entanglement peak and oscillation period. Namely, the generation of entanglement mainly depends on the weak measurement reversal in the weak coupling regime. But the big detuning can generate more entanglement than the weak measurement reversal in the strong coupling regime, and the greater the detuning is, the more entanglement is generated, and the entanglement time is also longer.

We provide physical explanations for the above results. Firstly, we explain the effect of cavity-environment coupling on entanglement dynamics. In the weak coupling regime, the negativity will reduce monotonously because the quantum information dissipates continuously into the environment, as illustrated in Fig.3a and 4a. In the strong coupling regime, the negativity will undergo damped oscillation due to the feedback and memory effects of environment, as depicted in Fig.3c and 4c. Secondly, we analyze the effect of SGI parameter on entanglement dynamics. The larger the SGI parameter is, the closer the two decay channels of each atom are parallel, and the faster the quantum information is exchanged between the atom and its cavity, thus the faster the negativity reduces for the maximal entangled state (see Fig.3), and the more entanglement is generated for the product state (see Fig.4). Thirdly, we discuss the effect of weak measurement and reversal on entanglement dynamics. From Eq. (36), we known that the bigger the weak measurement strength $p$ is, the more closely the initial qutrit is reversed towards the state $\left | 0 \right \rangle $, and the smaller the negativity is. If the qutrit is performed by the best weak measurement reversal, it can revert back to the excited state. Therefore, the bigger the weak measurement reversal strength is, the faster the negativity reduces for the maximal entangled state, and the more the entanglement is generated for the product state, as illustrated in Fig.2, 5 and 6. Finally, we consider the effect of detuning on entanglement dynamics. From Eq.(27), we know that, because the quantum information can been effectively trapped between two atoms by the detuning, the larger the detuning is, the slower the negativity reduces for the maximal entangled state , and the more the entanglement is generated for the product state, as illustrated in Fig.7 and 8.

\section{Conclusion}
In this paper we mainly studied how to modulate entanglement dynamics of two V-type atoms in dissipative cavity by detuning, weak measurement and reversal. First, we obtained the analytical solution of this model by solving Schr\"{o}dinger equation after diagonalizing Hamiltonian of dissipative cavity. Afterwards, we discussed in detail the influences of cavity-environment coupling, SGI parameter, detuning and weak measurement reversal on entanglement dynamics.  The results showed that the entanglement dynamics obviously depends on cavity-environment coupling, SGI parameter, detuning between cavity with environment and weak measurement reserval. For the maximal entangled state, the smaller SGI parameter, the stronger coupling, the bigger reversing measurement strength and the larger detuning can all protect the entanglement. The protection of entanglement mainly depends on the weak measurement reversal in the weak coupling regime. But in the strong coupling regime, the detuning protects the entanglement more effectively than the weak measurement reversal, which is more effectively than the SGI parameter. The maximal entanglement can be protected for a long time under the combination of strong coupling and the detuning , while the weak measurement reversal can make the entanglement arrive the steady value 0.82. For the product state, the entanglement can all be generated whether in the weak or strong coupling regimes. The entanglement generated obviously depends on an appropriate SGI parameter, a bigger reversing measurement strength and a larger detuning. The detuning can maximize the entanglement generated (close to 1.0) in the strong coupling regime, while the weak measurement reversal can make the entanglement generated arrive to the steady value 0.82. We also give the physical interpretations of all results.

\section*{Appendix}

We know Schr\"{o}dinger equation in the interaction picture is

\appendix
\setcounter{equation}{0}
\renewcommand{\theequation}{A\arabic{equation}}
\begin{equation}
i\frac{\mathrm{d}}{\mathrm{d}t}|\psi(t)\rangle=\hat{H}_{\mathrm{int}}(t)|\psi(t)\rangle
\end{equation}
where
\begin{equation}
\begin{split}
\hat{H}_{\mathrm{int}}(t)=\sum_{l=1}^2\sum_{m=A,B}\int\left(g_m\hat{\sigma}_m^{l+}\alpha^*(\omega)\hat{A}(\omega)e^{i(\omega_m-\omega)t}+H.C.\right)d\omega
\end{split}
\end{equation}

Let the total system has only one excitation and is initial in
\begin{equation}
|\psi(0)\rangle=(C_{2}^{A}\left(0\right)|C_{1}A_{2}\rangle+C_{1}^{B}\left(0\right)|B_{1}C_{2}\rangle)\otimes|0\rangle_{R}
\end{equation}
where $|C_{1}A_{2}\rangle$($|B_{1}C_{2}\rangle$) represents that the first(second) atom is in the ground state $|C\rangle$ and the other is in the excited state $|A\rangle$($|B\rangle$). $|C_{2}^{A}(0)|^{2}+|C_{1}^{B}(0)|^{2}=1$ and  $|0\rangle_{_R}=\hat{A}(\omega)|1_{\omega^{\prime}}\rangle\delta\left(\omega-\omega^{\prime}\right)$. $|0\rangle_{R}$ denotes that the environment is in the vacuum state. $|1_{\omega}\rangle$ indicates that the environment has only one excitation in the mode with frequency $\omega$.

We may write the time evolution state as
\begin{equation}
\begin{aligned}
|\psi(t)\rangle & =\left(C_{1}^{A}(t)|A_{1}C_{2}\rangle+C_{1}^{B}(t)|B_{1}C_{2}\rangle\right)_{S}\otimes|0\rangle_{\mathrm{R}}  \\
&\left.+\left(\left.C_{2}^{A}(t)\right|C_{1}A_{2}\right>+\left.C_{2}^{B}(t)\right|C_{1}B_{2}\rangle\right)_{S}\otimes|0\rangle_{\mathrm{R}} \\
&+\int C_{\omega}(t)|C_{1}C_{2}\rangle_{S}|1_{\omega}\rangle_{\mathrm{R}}\mathrm{d}\omega
\end{aligned}
\end{equation}

Substituting Eq.(A2) and (A4) into Eq.(A1), we can obtain the differential integrodifferentia  equation of the probability amplitude
\begin{equation}
\begin{aligned}
\dot{C}_{l}^{m}(t)=-ig_{m}\int\alpha^{*}(\omega)\mathrm{e}^{i(\omega_{m}-\omega)t}C_{\omega}(t)\mathrm{d}\omega \end{aligned}
\end{equation}
and
\begin{equation}
\begin{aligned}
\dot{C}_{\omega}(t)=-i\alpha(\omega)\sum_{m=A,B}g_{m}^{*}\mathrm{e}^{-i(\omega_{m}-\omega)t}\sum_{l=1}^{2}C_{l}^{m}(t)
\end{aligned}
\end{equation}

Substituting Eq.(A6) into Eq.(A5), we can obtain the differential integrodifferentia equation about ${C_{l}^{m}\left(t\right)}$
\begin{equation}
\begin{split}
\frac{dC_{l}^{m}\left(t\right)}{dt}=-\sum_{m,n=A,B}\int_{0}^{t}f_{mn}\left(t-t^{\prime}\right)\sum_{j=1}^{2}C_{j}^{n}\left(t^{\prime}\right)dt^{\prime}
\end{split}
\end{equation}
where
\begin{equation}
\begin{split}
f_{mn}(t-t')=\int d\omega J_{mn}(\omega)\mathrm{e}^{i(\omega_{m}-\omega)t-i(\omega_{n}-\omega)t'}
\end{split}
\end{equation}

Let that the environment has the Lorentzian spectral density as
\begin{equation}
J_{mn}(\omega)=\frac1{2\pi}\frac{\gamma_{mn}\kappa^2}{\left(\omega-\omega_c-\delta\right)^2+\kappa^2}
\end{equation}
where $\kappa$ is the spectral width of environment, and
$\gamma_{mn}=\frac{2g_mg_n^*}\kappa$  is the relaxation rate of the excited states, and
\begin{equation}
\gamma_{mm}=\gamma_m
\end{equation}
\begin{equation}
\gamma_{mn}=\sqrt{\gamma_m\gamma_n}\theta,m\neq n,|\theta|\le1
\end{equation}
where $\theta$ is defined as the  SGI (the spontaneously generated interference) parameter between the two decay channels $|A\rangle \to  |C\rangle$ and $|B\rangle \to |C\rangle$ of each atom, which $\theta$ depends on the angle between two dipole moments of the mentioned transitions.
\begin{equation}
\begin{split}
f_{AA}(t-t^{\prime})=f_{BB}\left(t-t^{\prime}\right)=f\left(t-t^{\prime}\right)  \\
=\int_{0}^{\infty}\mathrm{d}\omega J(\omega)\mathrm{e}^{i(\omega_{c}-\omega)\left(t-t^{\prime}\right)} \\
\end{split}
\end{equation}
\begin{equation}
\begin{split}
\begin{aligned}
f_{AB}(t-t^{\prime})& =f_{BA}\left(t-t^{\prime}\right)=f^{\prime}\left(t-t^{\prime}\right)  \\
&=\int_{0}^{\infty}\mathrm{d}\omega J^{\prime}\left(\omega\right)\mathrm{e}^{i\left(\omega_{c}-\omega\right)\left(t-t^{\prime}\right)}
\end{aligned}
\end{split}
\end{equation}
where$J^{\prime}(\omega)=\theta J(\omega)$.
Substituting Eq.(A9) into Eqs.(A12) and (A13), the following equations can be obtained respectively

\begin{equation}
\begin{aligned}
f(t-t')&=\frac{\gamma_0\kappa}{2}\mathrm{e}^{-( \kappa-i\delta )(t-t')}
\end{aligned}
\end{equation}

\begin{equation}
\begin{aligned}
f'(t-t')&=\frac{\gamma_0\theta\kappa}{2}\mathrm{e}^{-(\kappa-i\delta)(t-t')}
\end{aligned}
\end{equation}

Taking the Laplace transform from both sides of Eq.(A7), then we can get the following set of equations
\begin{equation}
\begin{split}
sC_{l}^{m}(s)-C_{l}^{m}(0)=-\sum_{n=A,B}\mathcal{L}\left\{f_{mn}(t)\right\}\sum_{j=1}^{2}C_{j}^{n}(s)
\end{split}
\end{equation}
where $C_l^m(s)=\mathcal{L}\{C_l^m(t)\}$ is the Laplace transform and
\begin{equation}
\begin{aligned}
\mathcal{L}\{f\left(t-t'\right)\}&=\frac{\gamma_0\kappa}{2\left(\kappa-i\delta +s\right)}\\\\\mathcal{L}\{f^{\prime}\left(t-t'\right)\}&=\frac{\gamma_0\theta\kappa}{2\left(\kappa-i\delta +s\right)}
\end{aligned}
\end{equation}

According to Eq.(A16), the following equation can be gotten
\begin{equation}
sC_l^m\left(s\right)-C_l^m\left(0\right)=sC_j^n\left(s\right)-C_j^n\left(0\right)
\end{equation}
Eq.(16) can be rewritten as
\begin{equation}
\begin{aligned}
&sC_l^m(s)-C_l^m(0) \\
&=-\sum_{n=A,B}\mathcal{L}\{f_{\mathrm{mn}}(t)\}  \\
&\times\left(2C_l^n(s)+\frac{1}{s}\sum_{j\neq l}^2\left(C_j^n(0)-C_l^n(0)\right.)\right)
\end{aligned}
\end{equation}
After defining the new coefficient
\begin{equation}
C_l^\pm(s)=C_l^A(s)\pm C_l^B(s)
\end{equation}

Substituting Eqs.(A14)-(A15) into Eq.(A19), we can rewrite Eq.(A19) as
\begin{equation}
\begin{aligned}
&sC_l^\pm(s)-C_l^\pm(0) \\
&=-\frac{\gamma_{0}(1\pm\theta)\kappa}{2(\kappa-i\delta +s)}  \\
&\times\left(2C_l^\pm(s)+\frac{1}{s}\sum_{j\neq l}^2\left(C_j^\pm(0)-C_l^\pm(0)\right)\right)
\end{aligned}
\end{equation}

Performing inverse Laplace transform on Eq.(A21), the following equation is
\begin{equation}
\begin{aligned}
&C_{l}^{\pm}\left(t\right)\\\\=&G_{\pm}\left(t\right)C_{l}^{\pm}\left(0\right)-\frac{\mathrm{1}-\mathrm{G}_{\pm}\left(t\right)}2\sum_{j\neq l}^{2}\left(C_{j}^{\pm}\left(0\right)-C_{l}^{\pm}\left(0\right)\right)
\end{aligned}
\end{equation}
where
\begin{equation}
\begin{aligned}
&G_\pm(t)=\mathrm{e}^{-(\kappa-i\delta ) t/2}\\&\times\left\{\cosh\left(\frac{d^\pm t}2\right)+\frac{\kappa-i\delta }{d^\pm}\sinh\left(\frac{d^\pm t}2\right)\right\}
\end{aligned}
\end{equation}
and
\begin{equation}
d^\pm=\sqrt{(\kappa-i\delta )^2-4\gamma_0\kappa(1\pm\theta)}
\end{equation}

According to Eq.(A20), we get the amplitudes
\begin{equation}
\begin{aligned}
C_l^A(t)&=\left(C_l^+(t)+C_l^-(t)\right)/2\\\\C_l^B(t)&=\left(C_l^+(t)-C_l^-(t)\right)/2
\end{aligned}
\end{equation}

\end{document}